\documentclass[journal]{IEEEtran}
\usepackage{cite}

\ifCLASSINFOpdf
\usepackage[pdftex]{graphicx}
\graphicspath{{pdf/}{jpeg/}{jpg/}}
\DeclareGraphicsExtensions{.pdf,.jpeg,.png,.jpg}
\else
\usepackage[dvips]{graphicx}
\graphicspath{{eps/}}
\DeclareGraphicsExtensions{.eps}
\fi
\usepackage{amsmath}
\usepackage{algorithmic}

\usepackage{array}
\usepackage{fixltx2e}

\usepackage{stfloats}
\usepackage{url}
\usepackage{xcolor}
\usepackage[T1]{fontenc}
\usepackage{amsmath}
\usepackage{amstext}
\usepackage{amssymb}
\usepackage{amsbsy}
\usepackage{amsopn}
\usepackage{amsfonts}
\usepackage{dsfont}
\usepackage{amsthm}
\usepackage{mathrsfs}
\usepackage{mathtools}
\usepackage[justification=centering]{caption}
\usepackage{graphicx}
\usepackage{natbib}

\newtheorem{thm}{Theorem}
\newtheorem{prop}{Proposition}

\newtheorem{assumption}{Assumption}
\newtheorem{remark}{Remark}

\begin{document}

	\title{Quadrotor going through a window and landing: An image-based visual servo control approach}
	\author{Zhiqi~Tang, Rita~Cunha, David~Cabecinhas, Tarek~Hamel, Carlos~Silvestre
	\thanks{This work was partially supported by the Project MYRG2018-00198-FST of the University
		of Macau; by the Macao Science and Technology, Development Fund under Grant FDCT/0031/2020/AFJ; by Funda\c{c}\~{a}o  para a Ci\^{e}ncia e a Tecnologia (FCT) through Project UIDB/50009/2020 and PTDC/EEI-AUT/31411/2017. The work of T. Hamel was supported by the DACAR project of the French National Research Agency (ANR). The work of Z. Tang was supported by FCT through Ph.D. Fellowship PD/BD/114431/2016 under the FCT-IST NetSys Doctoral Program.}
		
	\thanks{Z. Tang is with
		ISR, IST, Universidade de Lisboa, Portugal and I3S-CNRS, Universit\'{e} C\^{o}te d'Azur, Nice-Sophia Antipolis, France
		(zhiqitang@tecnico.ulisboa.pt).}
			\thanks{R. Cunha, D. Cabecinhas are with
		ISR, IST, Universidade de Lisboa, Portugal
		(rita@isr.tecnico.ulisboa.pt, dcabecinhas@isr.tecnico.ulisboa.pt).} \thanks{
		T. Hamel is with
		I3S-CNRS, Institut Universitaire de France, Nice-Sophia Antipolis, France
		(thamel@i3s.unice.fr).}
		\thanks{C. Silvestre is with
		Faculty of Science and Technology of the University of Macau, Macao, China and on leave from ISR, IST, Universidade de Lisboa, Portugal
		(csilvestre@umac.mo).}
}
\maketitle
%
\begin{abstract}
	This paper considers the problem of controlling a quadrotor to go through a window and land on a planar target, the landing pad, using an Image-Based Visual Servo (IBVS) controller that relies on sensing information from two on-board cameras and an IMU. The maneuver is divided into two stages: crossing the window and landing on the pad. For the first stage, a control law is proposed that guarantees that the vehicle will not collide with the wall containing the window and will go through the window with non-zero velocity along the direction orthogonal to the window, keeping at all times a safety distance with respect to the window edges. For the landing stage, the proposed control law ensures that the vehicle achieves a smooth touchdown, keeping at all time a positive height above the plane containing the landing pad. For control purposes, the centroid vectors provided by the combination of the spherical image measurements of a collection of landmarks (corners) for both the window and the landing pad are used as position measurement. The translational optical flow relative to the wall, window edges, and landing plane is used as velocity cue. To achieve the proposed objective, no direct measurements nor explicit estimate of position or velocity are required. Simulation and experimental results are provided to illustrate the performance of the presented controller.
\end{abstract}


%
\IEEEpeerreviewmaketitle

\section{Introduction}
\label{intro}
Navigation of Unmanned Aerial Vehicles (UAVs) using vision systems has been an important field of research during recent decades. Especially in indoor environments, where GPS is unavailable, a widely adopted alternative sensor suite includes an inertial measurement unit (IMU) and cameras, which are both passive, lightweight, and inexpensive sensors \cite{zingg10}. Three main solutions have been proposed for navigation using vision in indoor environments: map-based navigation, map-building-based navigation and mapless navigation \cite{desouza2002vision}. The first approach depends on a user-created geometric model or topology map of the environment, e.g. Perspective n Point (PnP), and the second requires the use of sensors to construct their own geometric or topological models, e.g. Simultaneous localization and mapping (SLAM). The mapless visual navigation, in which no global representation of the environment is required and the environment
is perceived as the system navigates, can be classified in accordance with the main vision technique or types of clues used during the navigation, which are methods based on optical flow, appearance and feature tracking (\cite{desouza2002vision}). However, the appearance-based method has the main problems which
are to find an appropriate algorithm for the representation of
the environment and to define the on-line matching criteria.

Visual servo control is a popular mapless navigation method based on feature tracking, which can be classified in two main categories: Image-based visual servo (IBVS) and Position-based visual servo (PBVS) control.  PBVS involves reconstruction of the
target pose with respect to the robot thus a 3-D model of the observed object should be known. However in IBVS, the control commands are deduced directly from image features, thus they offer advantages in robustness
to camera and target calibration errors, reduced computational
complexity, and simple extension to applications involving multiple
cameras compared to PBVS methods (\cite{hutchinson96}).
However, classical IBVS (\cite{chaumette2016visual}) suffers from three key problems. First, it is necessary to determine the depth of each visual feature used in the image error criterion independently from the control algorithm. Second, the rigid-body dynamics of the camera ego-motion are highly coupled when expressed as target motion in the image plane. And last, it uses a simple linearized control on the image kinematics that leads to complex non-linear dynamics and is not easily extended to the dynamics.

In order to overcome these problems, a spherical camera geometry can be used, from which the virtual spherical image points can be obtained by transforming the image points on the perspective camera to the view that would be seen by an ideal unified-spherical camera.
The passivity-like properties can be recovered for a centroid image feature as long as a spherical camera geometry is used. The novel IBVS algorithms based on spherical image centroids (e.g. applications on hovering an autonomous helicopter \cite{hamel02}, landing a quadrotor on moving platform \cite{herisse12,serra2016landing} and landing a fixed-wing aircraft on the runway \cite{lebras13,serra15,tang2018aircraft} ), do not require accurate depth information for observed image features and overcomes the difficulties associated with the highly coupled dynamics of the camera ego-motion in the image dynamics.

\begin{figure}[!htb]
	\centering
	\includegraphics[scale = 0.5]{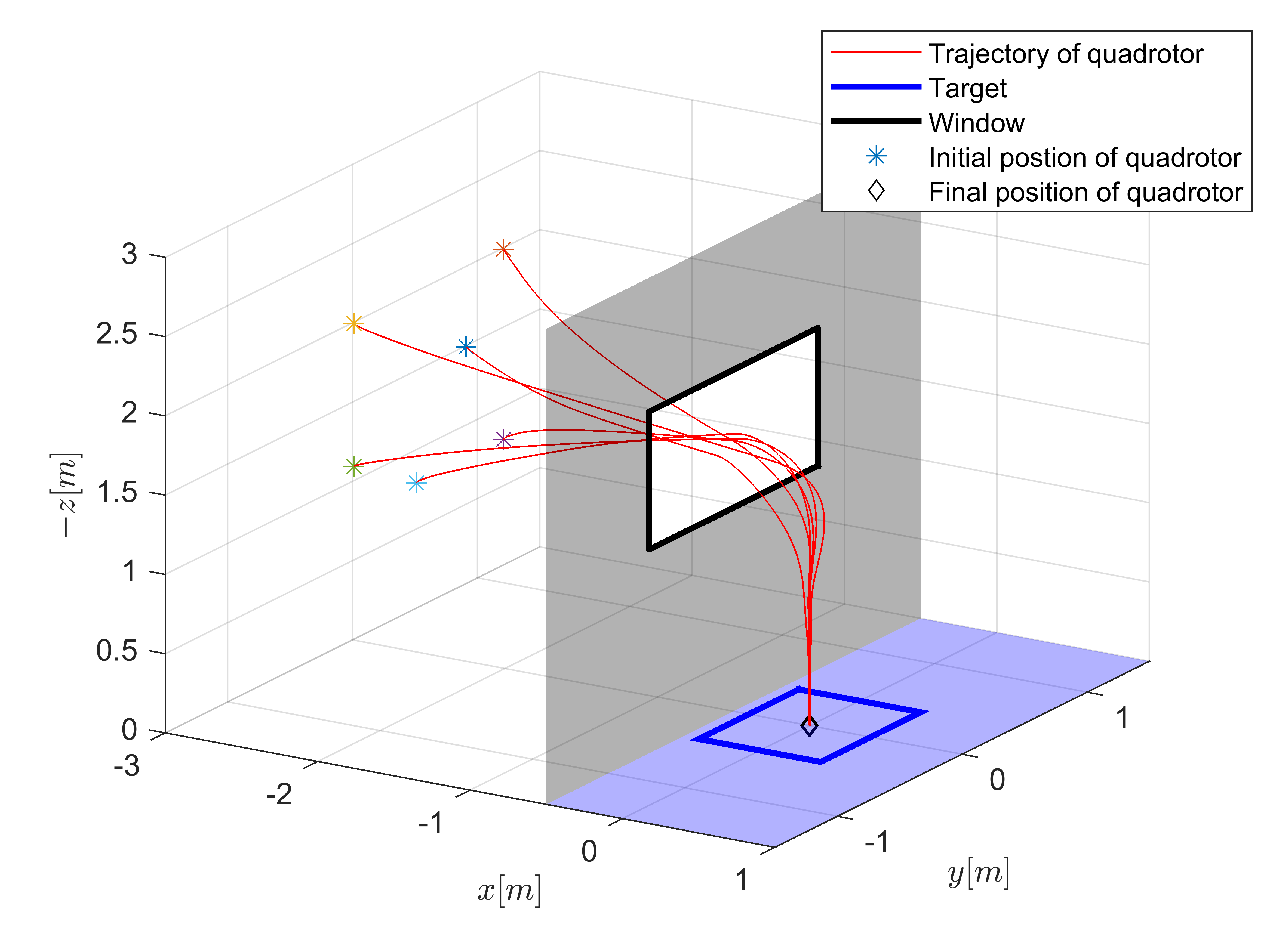}
	\caption{3-D plot of the quadrotor trajectories under different initial conditions.}
	\label{fig:3D}
\end{figure}

This paper extends the IBVS control solution based on spherical image centroids to a specific problem of steering a quadrotor to move from one room to a second one by crossing a window and then land on a planar target placed in the second room (see Fig.~\ref{fig:3D}). 
This application has significant practical interest since many tasks (i.e. search and rescue in an earthquake-damaged building \cite{michael2012collaborative}, package delivery using UAVs) require UAVs to land on a final destination or to perform intermediate landings for battery recharge or exchange, or refueling (for larger UAVs) during long missions. The quadrotor is assumed to be equipped with an IMU and two on-board cameras: one forward-looking 
and another downward-looking. 
Neither the translational velocity and position of the vehicle nor the location of the target (window and landing pad) are known. In the proposed IBVS control laws, the centroid vectors provided by the combination of the spherical image measurements of a collection of landmarks (corners) from both the window and the landing pad are used as position cue and the translational optical flow relative to the plane containing window and landing pad is used as velocity cue.

The proposed control strategy draws inspiration from \cite{serra2016landing} which combines a centroid-like feature and the translational optical flow to perform exponential landing on a desired spot. This paper considers different control objectives: going through a window and then landing on a desired target. The control law for going through a window ensures that no collision with the wall or windows edges will occur and the vehicle will align with the center line orthogonal to the window, crossing it with non-zero velocity. The control law for the landing is an improvement with respect to the one used in \cite{serra2016landing} in which the desired optical flow was chosen constant, leading to a high-gain controller that fails to achieve a perfect landing maneuver. Conversely, the desired optical flow adopted in this work is not constant. It corresponds to the component of the image centroid in the direction orthogonal to the target plane leading to a vanishing desired optical flow when the distance to the target approaches zero and therefore one avoids the high-gain nature of prior work (\cite{serra2016landing}).


Following on previous work \cite{tang2018going}, which presents the preliminary results with only simulations, this paper presents the following novel contributions: 1) bounded disturbances
(e.g. due to wind, and unmodeled dynamics) are included in the dynamics of the system; 2) a complete stability analysis shows that convergence to the desired zero-height equilibrium is guaranteed in all cases and ultimate boundedness of the horizontal position error is guaranteed when landing in the presence of horizontal disturbances; 3) experimental results are provided where the controllers run on an onboard computer together with the image processing for the detection of window and landing pad and for the computation of the translational optical flow.


The body of the paper consists of eight parts. Section \ref{sec:model} presents the dynamic model, the fundamental equations of motion, and the adopted hierarchical control architecture. Section \ref{sec:image feature} introduces the environment and presents the image features that are used in the control laws. Section \ref{sec:controller} proposes two control laws: one for the landing task in obstacle-free environments and the other for flying through the window. A combination of these two control laws in the practical case is also presented in this section. Section \ref{sec:simu} shows simulation results obtained with the proposed controller. Section \ref{sec:experiment} presents and analyzes the experimental results which validate the proposed controllers. The paper concludes with some final comments in Section \ref{sec:conclusion}.\\

\subsection{Related Work}\label{sec:related work}

There are several examples in the literature of recent work dedicated to the problem of flying autonomous vehicles in complex environment using vision systems. In \cite{loianno2017estimation,falanga2017aggressive,guo2020image}, the authors specifically address the problem of going through a window using only a single camera and an IMU. However, estimation of vehicle's position and velocity is required in \cite{loianno2017estimation,falanga2017aggressive}. Besides, the pose of the window is assumed to be known in \cite{loianno2017estimation}. Although the work in \cite{guo2020image} directly uses image feature as position cue, estimates of the image depth are still required and the velocity vector is assumed to be known.  In general, state estimation adds computational complexity, and the output is often sensitive to image noise and camera calibration errors. The limited work on image-based control approach can be explained by the complexity involved in obtaining sound proofs of convergence and stability.

Landing in complex environments calls for obstacle avoidance capabilities, which are naturally provided by the use of optical flow, a visual feature that draws inspiration from flying insects. Optical flow measures the pattern of apparent motion of objects, surfaces, and edges in a visual scene caused by the relative motion between an observer and a scene (\cite{burton1978thinking}). It has been experimentally shown that the neural system of the insects reacts to
optic flow patterns to produce a large variety of flight capabilities, such
as obstacle avoidance, speed maintenance, odometry estimation, wall following and corridor centering, altitude regulation, orientation control
and landing (\cite{floreano2015science,serres2017optic}). 
Using optical flow as velocity cue and observed feature expressed in terms of an unnormalized
spherical centroid, a fully nonlinear adaptive visual servo control design is provided in \cite{mahony08}. Although estimating the height of the camera above the landing plane was still required, it was the first time that an IBVS control using image measurements for both position and velocity was proposed, going beyond the kinematic model to consider the dynamics. Based on \cite{mahony08} and using optical flow, the authors proposed IBVS controllers for landing a quadrotor (\cite{herisse12,serra2016landing}) and landing a fixed-wing aircraft eliminating the need to estimate the height of the vehicle above the ground (\cite{lebras13,serra15,tang2018aircraft}). Using a distinct paradigm, a novel setup of self-supervised learning based on optical flow was introduced in \cite{ho2018optical}. Using optical flow, the proposed method learns the visual appearance of obstacles in order to search for a landing spot for micro aerial vehicles.

When compared to related work, this paper proposes simple IBVS controllers applied in sequence to first go through a window and then land on a planar target, using only vision measurements and requiring no estimation of position, velocity, image depth, nor height above the target. The present work also provides rigorous mathematical proofs for stability and robustness in the presence of disturbances, complemented by experimental validation of the proposed controllers.


\section{Quadrotor Modeling and control architecture} \label{sec:model}
\begin{figure}[!t]
	\centering
	\includegraphics[width=2.8in]{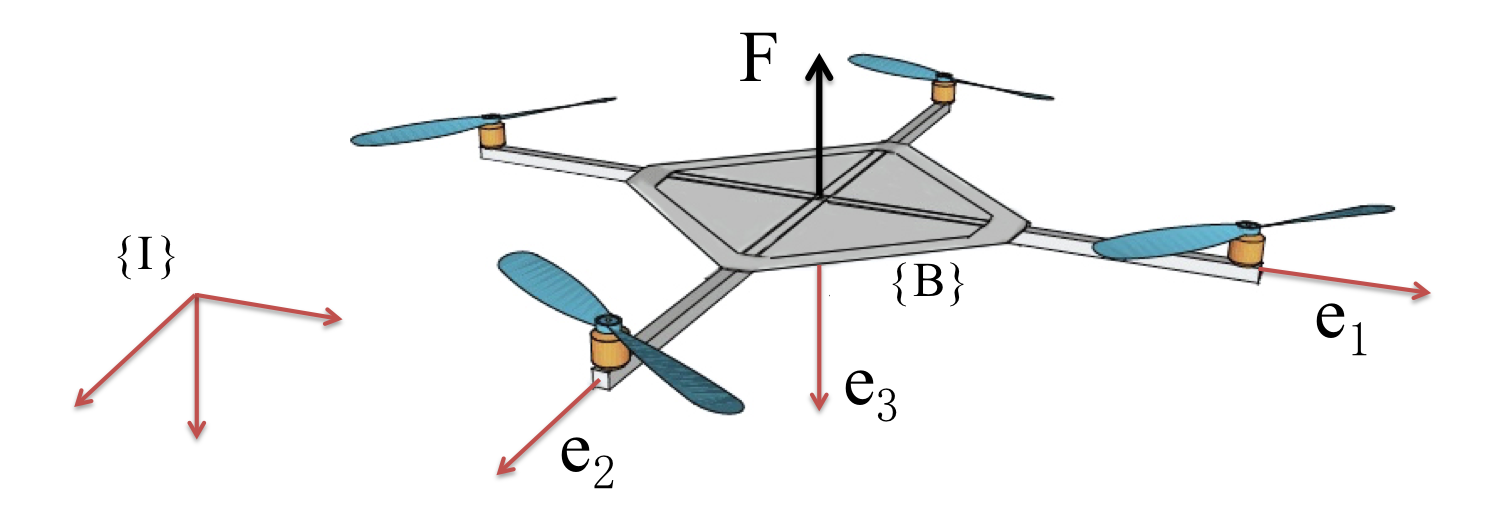}	
	\caption{Reference frames and force for schematic representation of a
		quadrotor.}
	\label{fig:vehicle}
\end{figure}
Consider a quadrotor equipped with an IMU and two cameras. To describe the motion of the quadrotor, two reference frames are introduced: an inertial reference frame \{$I$\} fixed to the earth surface and a body-fixed frame \{$B$\} attached to the quadrotor's center of mass (see Fig. \ref{fig:vehicle}). Let $R={_{B}^{I}}R\in SO(3)$ denote the orientation of the frame \{$B$\} with respect to \{$I$\} and let $\xi\in \mathbb R^3$ be the position of the origin of the frame \{$B$\} with respect to \{$I$\}. Let $v\in \mathbb R^3$ denote the translational  velocity expressed in \{$I$\} and $\Omega\in \mathbb R^3$ the orientation velocity expressed in \{$B$\}. The kinematics and dynamics of the quadrotor vehicle are then described as
\begin{equation}
\left \{
\begin{aligned}
\dot\xi&=v\\
m\dot v&=-F+mge_3+\triangle \label{eq:system}
\end{aligned}
\right.
\end{equation}
\begin{equation}
\left \{
\begin{aligned}
\dot{R}&=RS(\Omega) \label{eq:R}\\
\bold{I}\dot \Omega&=-\Omega_{\times}\bold{I}\Omega+\Gamma
\end{aligned}
\right.
\end{equation}
with $g$ the gravitational acceleration, $m$ the mass of the vehicle and $\bold{I}$ its inertia matrix. The matrix $\Omega_{\times}$ denotes the skew-symmetric matrix matrix associated with the vector product $\Omega_{\times}x=\Omega \times x$, for any $x\in \mathbb{R}^3$.

The vector $F \in \mathbb{R}^3$ expressed in \{$I$\} combines the principal non-conservative forces applied to the quadrotor and generated by the four rotors. In quasi-hover conditions one can reasonably assume that this aerodynamic force is always in the direction $e_3^b$ in \{$B$\}, since all the four thrusters  are aligned with $e_3^b $ and their contribution predominates over other components.
Thus the $F$ in the direction of $e_3^b$ expressed in the inertial frame can be described as follows:
\begin{equation}
F=F_TR e_3 \label{eq:F}
\end{equation}
where the scalar $F_T$ represents the total thrust magnitude generated by the four motors. It also represents the unique control input for the translational dynamics.

The term $\triangle$ combines the modeling errors and aerodynamic effects due to the interaction of the rotors wake with the environment  causing random wind and dynamic inflow effects (\cite{peters1988dynamic}).

The vector $\Gamma \in \mathbb{R}^3$ expressed in \{$B$\} is the torque control for the attitude
dynamics. It is obtained via the combination of the contributions of four rotors. The invertible linear map between $[F_T \in \mathbb{R}^+, \Gamma \in \mathbb{R}^3]$ and the collection of individual thrusters $[F_{T_1},F_{T_2},F_{T_3}, F_{T_4}]$ can be found in \cite{hamel2002dynamic}.

\begin{figure}[!t]
	\centering
	\includegraphics[width=3.4in]{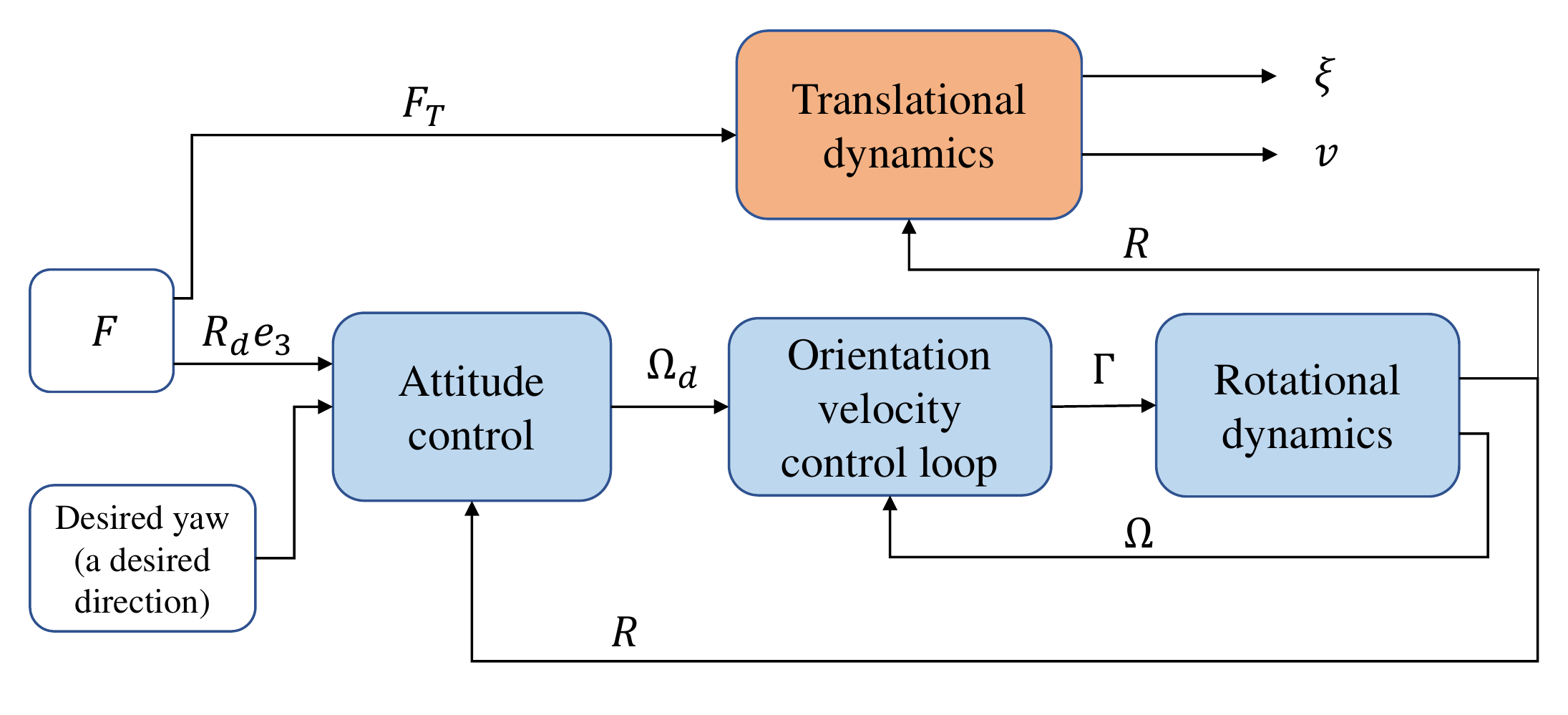}	
	\caption{A hierarchical control design strategy.}
	\label{fig:hiearchical}
\end{figure}
\subsection{Control architecture}
A hierarchical control design strategy is adopted in this paper (see Fig. \ref{fig:hiearchical}). This choice is motivated by the natural structure of the system dynamics and its practical implementation (\cite{bertrand11,herisse12}). For the translational dynamics of the quadrotor (Eq. \eqref{eq:system}), the force $F$ (Eq. \eqref{eq:F}) is used as control input  by means of its thrust direction  and its  magnitude.  This constitutes a high-level outer loop for the control design. The thrust $F_T$ is directly the magnitude of the designed force ($F_T=\|F\|$) and the desired attitude $R_d$ (partly obtained by the desired direction $R_d e_3=\frac{F}{\|F\|}$ complemented by a desired yaw) can then be reached by considering the body's angular velocity $\Omega$  as an intermediary control input, which constitutes again a desired angular velocity for the fully actuated orientation dynamics (Eq. \eqref{eq:R}) via the high gain control torque $\Gamma$. The stabilisation of
the orientation dynamics is not the subject of this paper and it is assumed that a suitable low level robust stabilising control is implemented, that satisfactorily regulates the attitude error with a fast dynamics.

\section{Environment and Image Features}\label{sec:image feature}
In this section adequate image features in relation to the considered tasks are derived and all required assumptions regarding the environment and the setup are established.
\begin{figure}[!t]
	\centering
	\includegraphics[width=3.5in]{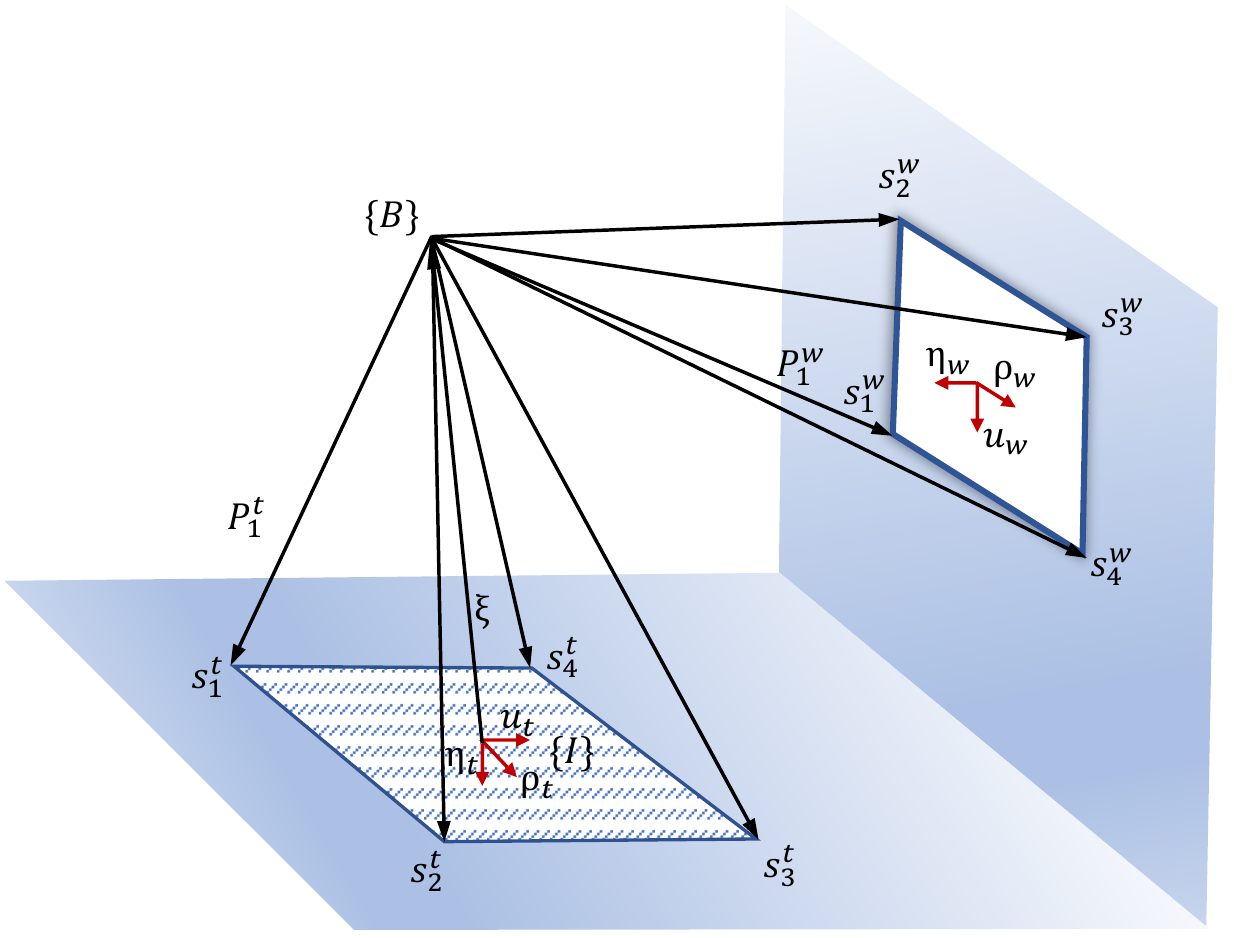}	
	\caption{Landing plane and window plane.}
	\label{fig:enviro}
\end{figure}
\begin{assumption}
	A downward-looking camera and a forward-looking camera are attached to the center of mass of the vehicle. The downward-looking camera reference frame coincides with the body-fixed frame \{$B$\}. The rotation matrix from the forward-looking camera reference frame to the body frame ${_{C}^{B}}R\in SO(3)$ is known. \label{assum:1}
\end{assumption}
\begin{assumption}\label{ass:IMU}
The angular velocity $\Omega$ is measured and the orientation matrix $R$ of \{$B$\} with respect to \{$I$\} is obtained by external observer-based IMU measurements.  This allows to represent all image information and the system dynamics in the inertial frame.
\end{assumption}
\begin{assumption}\label{ass:landing}
	The landing target lies on a textured plane which is called target plane. Its normal direction $\eta_t\in \mathbb S^2$ in the inertial frame is known (typically $\eta_t \approx e_3$).
\end{assumption}	

\begin{assumption}
	The target window has a rectangle shape and lies on a textured wall which is called window plane. Its width $r_w$ is known but its normal direction $\eta_w\in \mathbb S^2$ is unknown.
\end{assumption}

Both landing plane and window plane are placed in the environment, as shown in Figure \ref{fig:enviro}. It assumed that the vehicle is able to recognize the landing pad and the window from landmarks on the pad and from corners and edges of the window respectively. The background texture on both landing plane and window plane are also exploited to obtain information about the vehicle's velocity with respect to the planes and also to avoid collisions with the wall and the window's edges.

For any initial position (along with any initial velocity) outside the room containing the landing pad, the main objective is to design a feedback controller resorting only to image features that can ensure automatic landing of the vehicle without any collision.
\subsection{Image features on the landing plane}\label{sec:image for target}
The target on the landing plane is depicted in Figure \ref{fig:enviro}. The axes of $\{I\}$ are given by $(u_t,\rho_t,\eta_t)$, where $\rho_t=\eta_t\times u_t$, and the origin of $\{I\}$ is placed at the center of the landing pad. As shown in Figure \ref{fig:enviro}, $s_{i}^t\in \mathbb R^3$ denotes the position of $i$th marker (or a corner) of the landing pad relative to the inertial frame expressed in $\{I\}$. Note that $\eta_t^\top s_i^t=0$. 
Define the position vector of $i$th marker of the target relative to $\{B\}$ as
\begin{equation}
P_{i}^t=s_{i}^t-\xi.
\end{equation}
The position of the vehicle relative to the center of the landing pad is defined as
\begin{equation}\label{eq:xit}
\xi_t=-\frac 1 {n_t}\sum_{i=1}^{n_t}P_{i}^t=\xi -\frac 1 {n_t}\sum_{i=1}^{n_t}s_{i}^t
\end{equation}
where $n_t$ is the number of observed markers on the landing pad and $\frac 1 {n_t}\sum_{i=1}^{n_t}s_{i}^t$ is a constant vector. This sum is zero when all markers are in the camera field of view.

Using the spherical projection model for a calibrated camera, the spherical image points of landing pad's markers can be expressed as
\begin{equation}
p_{i}^t=\frac {P_{i}^t} {\|P_{i}^t\|}=\frac{s_{i}^t-\xi}{\|s_{i}^t-\xi\|} \label{ptt}
\end{equation}
It can be obtained from the 2D pixel locations $(X_i^t,Y_i^t)$ of the camera image, such that
\begin{align}
p_i^t=R\frac{\bar p_i^t}{||\bar p_i^t||} ,\;\mbox{ with }\;\bar p_i^t=A^{-1}
\begin{bmatrix}
X_i^t\\Y_i^t\\1
\end{bmatrix}.\label{eq:barpi}
\end{align}
The matrix $A^{-1}$ in the above equation is the camera's intrinsic parameters that transforms image pixel to perspective coordinates $\bar p_i^t$. Note that expressions \eqref{ptt} and \eqref{eq:barpi} are the same and hence $p_{i}^t$ does not depend on the orientation.

The visual feature used for the landing task is the the centroid of the observed visual feature.
\begin{align}
q_t:=-\frac 1 {n_t} \sum_{i=1}^{n_t}p_{i}^t= -R\left(\frac 1 {n_t}\sum_{i=1}^{n_t}\frac{\bar p_i^t}{||\bar p_i^t||}\right)\label{eq:q_t}
\end{align}
which is the simplest  image feature that encodes all information about the position of the vehicle with respect to the landing plane. 
It is not necessary to match observed image points with desired features as required in classical image based visual servo control. Besides, the calculation of the image centroid is highly robust to pixel noise, and easily computed in real-time in the camera frame and then derotated. This ensures that $q_t$ is invariant to any orientation motion (\cite{hamel02}).


\subsection{Image features on the window plane}
As shown in the Figure \ref{fig:enviro}, a rectangular window is placed on a textured wall. Its corners and edges are assumed to be recognized in camera images. Both information are combined together to extract the normal direction $\eta_w$ and provide the feedback information used in the controller.

Consider first the the windows corners and let $s_{i}^w\in \mathbb R^3$ denote the position of $i$th corner of the window expressed in $\{I\}$. Define the position vector of $i$th corner of the window relative to $\{B\}$ as
\begin{equation}
P_{i}^w=s_{i}^w-\xi. \label{P_w}
\end{equation}
From there, one can deduce the position of the vehicle with respect to the window's center:
\begin{equation}
\xi_w=-\frac 1 {n_w} \sum_{i=1}^{n_w}P_{i}^w=\xi-\frac 1 {n_w}\sum_{i=1}^{n_w}s_i^w, \label{eq:chi}
\end{equation}
with $n_w$ (typically $n_w=4$) number of the window's corners and $\frac 1 {n_w}\sum_{i=1}^{n_w}s_i^w$ constant vector.

Similarly to Section \ref{sec:image for target} and recalling that the forward-looking camera is used to detect the window, the spherical image points of the corners of the window are exploited:
\begin{align}
p_{i}^w=\frac {P_{i}^w} {\|P_{i}^w\|}= R\, {_{C}^{B}}R \frac{\bar p_i^w}{||\bar p_i^w||}, \label{eq:pwi}
\end{align}
with $\bar p_i^w$ the perspective coordinates of the $i$th window's corner, leading the following centroid:
\begin{equation}
q_w(t):=-\frac 1 {n_w} \sum_{i=1}^{n_w}p_i^w(t)=- R\, {_{C}^{B}}R \left(\frac 1 {n_w} \sum_{i=1}^{n_w}\frac{\bar p_i^w}{||\bar p_i^w||}\right) \label{q_w},
\end{equation}
where ${_{C}^{B}}R$ is the rotation matrix from the forward-looking camera reference frame to the body frame.

\begin{figure}[!t]
	\centering
	\includegraphics[width=2in]{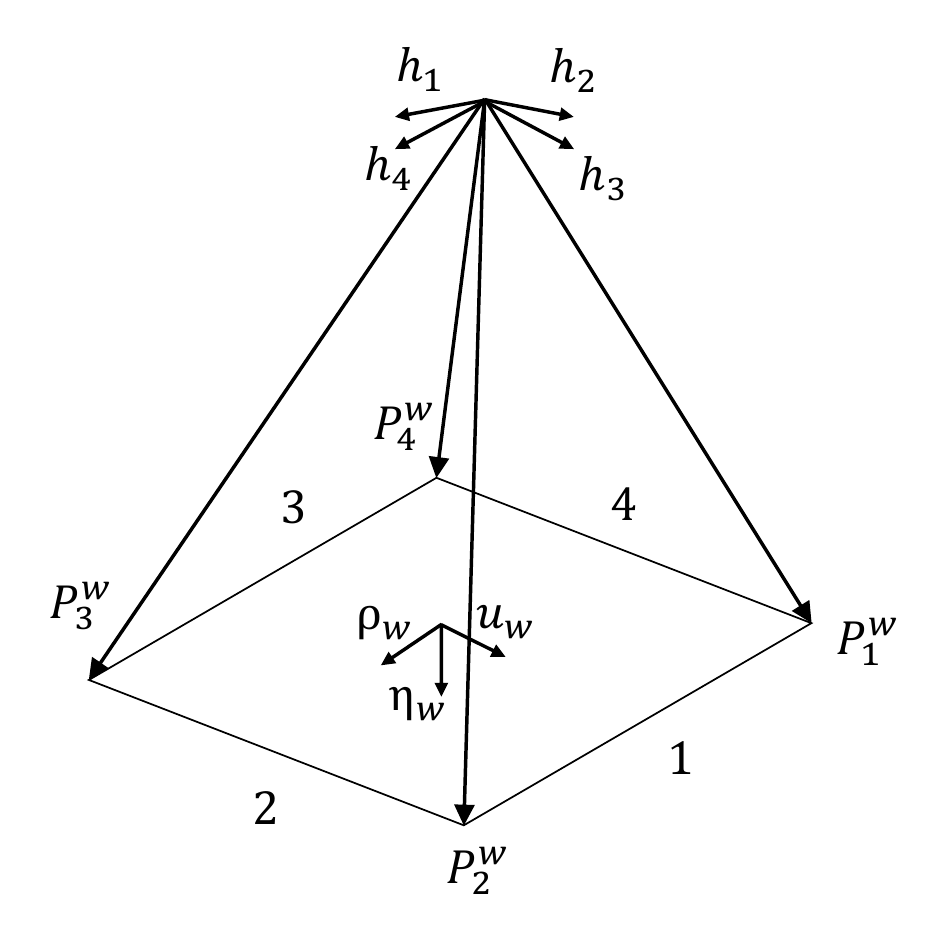}
	\caption{Window plane and unit directions $h_i$ normal to the planes defined by the origin of camera frame and the $i$th window edge.}
	\label{fig:window}
\end{figure}

Now, to extract the normal direction $\eta_w$, recall that the axes representing the window are given by $(\eta_w,\rho_w,u_w)$, with $\rho_w=u_w\times \eta_w$ (see Figure \ref{fig:window}). Using the image of $i$th line and exploiting the fact that the window has a rectangular shape, it is straightforward to get the directions $u_w$ and $\rho_w$ and consequently $\eta_w$.
As described in \cite{mahony05}, in the binormalized Euclidean Plucker coordinates, the $i$th line can be represented by its unit direction $u_w$ (resp. $\rho_w$) and the unit direction $h_i$, which is normal to the plane defined by the origin of the camera/body-fixed frame and the $i$th line. The unit vector $h_i$ can be obtained directly from the images of lines which can be identified using a convenient line detection technique, such as the Hough transform. Using the fact that lines $1$ and $3$ (resp. lines $2$ and $4$) are parallel in the inertial frame, one deduces the measure of the direction $u_w$ (resp. $\rho_w$) from the following relationships:
\begin{align}
\rho_w&=\pm\frac{h_1\times h_3}{\|h_1\times h_3\|}\\
u_w&=\pm\frac{h_2\times h_4}{\|h_2\times h_4\|}.
\end{align}
Then the normal vector to the window plane is directly obtained by
\begin{align}\label{eq:etaw}
\eta_w=\pm\frac{u_w\times \rho_w}{\|u_w\times \rho_w\|}
\end{align}
and the sign of equation \eqref{eq:etaw} is chosen such that the condition $\eta_w^\top q_w(0)<0$, with $q_w(t)$ the image centroid of window's corners in equation \eqref{q_w}.

To exploit the image of window's edges, defining the vector from the vehicle to the closest point on window's edges as $L^e\in \mathbb R^3$, its direction $l^e=\frac{L^e}{\|L^e\|}$ can be obtained from the camera
\begin{align}
l^e=\{l_i^e:\max\{|\eta_w^\top l_i^e|\},i=\{1,2,3,4\}\}
\end{align}
where
\[l_i^e=\pm (h_i \times \rho_w),  i=\{1,3\}, \;\; l_i^e=\pm (h_i \times u_w), i=\{2,4\} \]
are the directions from the vehicle to the nearest point on each edge $i$.

\begin{figure}[!t]
	\centering
	\includegraphics[width=2.5in]{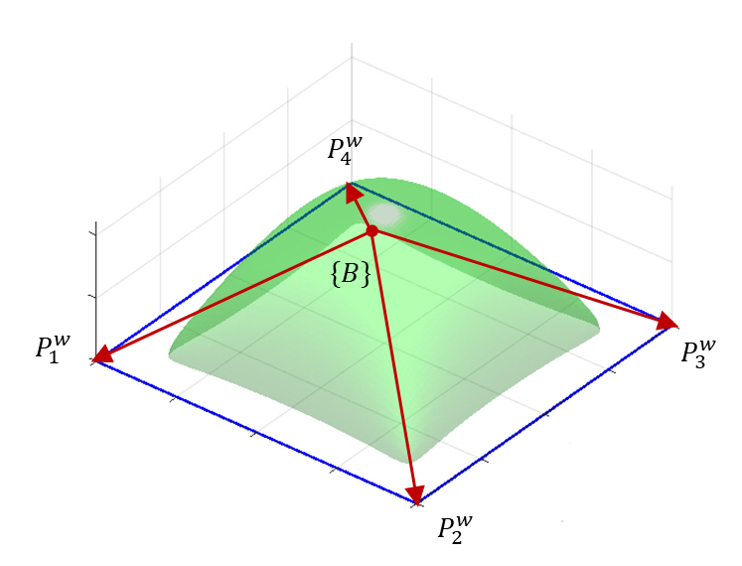}
	\caption{The green volume represents the region $\mathcal W$ defined by inequality \eqref{eq:region} which excludes the window edges.}
	\label{fig:window_area}
\end{figure}

Form now on, it is able to derive the required information achieving the double goal of going through the window in the meanwhile avoiding collision with the window edges and wall. We first define the safety region $\mathcal{W}$ such that
\begin{align}\label{eq:region}
\mathcal{W}&:=\{\xi_w:\|q_w(\xi_w)\|\le \epsilon\},
\end{align}
where $\epsilon > 0$ is chosen such that $\forall\xi_w\in\mathcal{W}$, the condition $\|\xi_w\|<\frac {r_w}2-\epsilon$ also holds, implying that the region $\mathcal{W}$ does not contain the window edges (see Fig. \ref{fig:window_area}).

From there, the chosen visual feature that encodes all required information about the position of the vehicle with respect to the window is:
\begin{equation}\scalebox{0.92}{$
	\bar{q}_w:=
	-\frac 1 {n_w} \sum_{i=1}^{{n_w}} p_i^w \left[\alpha_w(t)\frac{1}{\eta_w^\top p_i^w}+(1-\alpha_w(t))\frac{\eta_w^\top  l_e}{\eta_w^\top p_i^w}\right], \label{eq:qw}$}
\end{equation}
where $\alpha_w(\|q_w(\xi_w)\|)$ is a weight function ensuring the continuity of $\bar{q}_w$. It is defined as follows:
\begin{equation}
\alpha_w(t)=
\begin{cases}
0 &\text{, if } \|q_w\| \le \epsilon\ (\xi_w \in \mathcal W)\\
\frac{1}{\delta}(\|q_w\|-\epsilon), & \text{, if} \;\epsilon<\|q_w\|< \epsilon+\delta\\
1&\text{, if } \|q_w\|\ge \epsilon+\delta,
\end{cases}\label{eq:alphaw}
\end{equation}
with $\delta$ an arbitrary small positive constant. Since $\eta_w^\top l^e = \frac{\eta_w^\top L^e}{\|L^e\|}=\frac{\eta_w^\top P_i^w}{\|L^e\|}$, $\bar{q}_w$ can be expressed in terms of the unknown distance $d_o$ and $d_e$:
\begin{equation}\label{d_w}
\bar{q}_w(t)= \alpha_w(t)\frac {\xi_w(t)} {d_o(t)} + (1-\alpha_w(t)) \frac {\xi_w(t)} {d_e(t)}
\end{equation}
where $d_o:=\eta_w^\top P_i^w=-\eta_w^\top\xi_w$ is the distance from the camera to the wall and  $d_e:=\|L^e\|=\sqrt{d_o^2+\|\pi_{\eta_w}L^e\|^2}$ represents the distance from the camera to the closest window's edge.
\subsection{Image Kinematics and Translational Optical Flow}
The kinematics of any observed points on the landing plane (including markers of the the landing pad) can be written as:
\begin{equation}
\dot P^t=-\dot{\xi}=-v
\end{equation}
where $P^t$ expressed in $\{I\}$ denotes any point on the textured ground of the landing plane.
So the kinematics of the corresponding image point $p^t=\frac {P^t} {\|P^t\|}$ can be expressed as
\begin{equation}
\dot p^t=-\pi_{p^t}\frac v {\|P^t\|}. \label{eq:dotp}
\end{equation}
with
\begin{align*}
\pi_y := I_3 - y y^{\top} \geq 0, \label{eq:pi}
\end{align*}
the orthogonal projection operator in $\mathbb{R}^3$ onto the $2$-dimensional vector subspace orthogonal to any $y \in \mathbb{S}^2$.
Let $d_t$ be the height of the vehicle above the landing plane:
\begin{equation}
d_t:=\eta_t^\top P^t=\eta_t^\top P^t_i=-\eta_t^\top \xi_t,
\end{equation}
then equation (\ref{eq:dotp}) can be rewritten as
\begin{equation}\label{p_t}
\dot p^t=-\cos\theta^t\pi_{p^t} \phi_t(t)
\end{equation}
where $\cos\theta^t=\frac{d_t}{\|P^t\|}=\eta_t^\top p^t$ and $\phi_t$ is the translational optical flow:
\begin{align}
\phi_t(t)=\frac{v(t)}{d_t(t)} \label{eq:opticalT}
\end{align}
which is the ideal image velocity cue that can be complemented with the centroid information for designing a pure IBVS controller to perform the landing task. 

The translational optical flow $\phi_t$ can be obtained by integrating $\dot p^t$ \eqref{p_t} over a solid angle $S^2$ of the sphere around the normal direction $\eta_t$ to the landing plane. It can be shown that the average of the optical flow is calculated as in \cite{herisse12}:

\begin{equation}
\phi_t(t)=-(R_t\Lambda^{-1}R_t^\top)\int \int_{S^2} \dot p^t dp^t \label{p_ti},
\end{equation}
where matrix $\Lambda$ is a constant diagonal matrix depending on parameters of the solid angle $S^2$, and $R_t$ represents the orientation matrix of the landing plane with respect to the inertial frame. Since \{$I$\} is chosen coincident with the target frame one has $R_t=I_3$.

In practice, the optical flow is first measured in the camera frame from the 2-D optical flow $\dot {\bar p}_t$ obtained from a sequence of images using the Lucas-Kanade algorithm and then derotated (see \cite{herisse12} for more detail).  Note however that computing the optical flow from \eqref{p_ti} or directly from $\dot {\bar p}_t$ in the camera frame and then derotating it, the result is theoretically the same and does not depend on the measured $\Omega$ nor on the estimated $R$.


Similarly, the kinematics of any observed points on the window plane can be written in the inertial frame as
\begin{equation}
\dot P^w=-v
\end{equation}
where $P^w$ expressed in $\{I\}$ denotes the position of a point on the textured wall of the window plane with respect to \{$B$\} expressed in \{$I$\}, not to be confused with $P_{i}^w$ in Eq. \eqref{P_w}, which is the position of the $i$th corner of the window with respect to \{$B$\} and also expressed in \{$I$\}. So the kinematics of the corresponding image point $p^w=\frac {P^w} {\|P^w\|}$ can be written as
\begin{equation}
\dot p^w=-\cos\theta^w\pi_{p^w}\frac v {d_o}
\end{equation}
with $\cos\theta^w=\frac{d_o}{\|P^w\|}=\eta_w^\top p^w$.
Analogously to the previous case, the translational optical flow with respect to the textured wall $\frac v {d_o}$ can be obtained from the integral of $\dot p^w$ along the direction $\eta_w$ over a solid angle.

Now, to achieve the goal that the vehicle is going through the window smoothly, the translational optical flow with respect to the closest window's edge is also used. The kinematics of any observed points on the closest window's edge is
\begin{equation}
\dot P^{e}=-v
\end{equation}
where $P^{e}$ denotes the position of a point on the the closest edge from the window. The kinematics of the corresponding image point $p^e=\frac {P^{e}} {\|P^{e}\|}$ can be written as
\begin{equation}
\dot p^e=-\cos\theta^e\pi_{l^e}\frac v {d_e}
\end{equation}
with $\cos\theta^e=\frac{d_e}{\|P^{e}\|}=l^{e\top} p^e$. The translational optical flow with respect to the closest window edge, $\frac v {d_e}$, can be obtained from the integral of $\dot p^e$ along the direction $l^e$ over a solid angle.

Analogously to \eqref{eq:qw}, the translational optical flow used for going through the window is the convex combination of the translational optical flow with respect to the textured wall and to the closest window edge, respectively:
\begin{equation}\label{eq:phi_w}
\phi_{w}= \alpha_w(t)\frac {v(t)} {d_o(t)} +  (1-\alpha_w(t))\frac {v(t)} {d_e(t)}
\end{equation}
with $\alpha_w(t)$ defined already by \eqref{eq:alphaw}.
\section{Controller design}\label{sec:controller}
\subsection{Landing in obstacle free environment}
\begin{thm}\label{th:landing}
	Consider the system (\ref{eq:system}) in the nominal case ($\triangle \equiv 0$) subjected to the following feedback control:
	\begin{align}
	F_t=K^t_pq_t+K^t_d\phi_t+mge_3. \label{eq:FT}
	\end{align}
	with $K_p^t=k_{p_{1,2}}^t \pi_{\eta_t}+k_{p_3}^t \eta_t \eta_t^\top$ and $K_d^t=k_{d_{1,2}}^t \pi_{\eta_t}+k_{d_3}^t \eta_t \eta_t^\top$ two constant positive definite matrices.
	If for any initial condition such that $d_t(0)=-\eta_t^\top \xi_t(0) \in \mathbb{R}^+$, then the following assertions hold $\forall t\ge 0$:
	\begin{enumerate}
	\item the height $d_t(t)=-\eta_t^\top \xi_t(t)\in \mathbb{R}^+$ and its derivative $\dot{d}_t(t) \in \mathbb{R}$ are well defined and uniformly bounded $\forall t$ and converge to zero asymptotically,
	\item the acceleration $\dot{v}(t)$ and the states $(\xi_t(t),v(t))$ are bounded and converge asymptotically to zero.
	\end{enumerate}
\end{thm}
\begin{proof}
	See Appendix A.
\end{proof}

\begin{prop}\label{prop:landing}
	Consider the system (\ref{eq:system}) in which $\triangle$ and $\dot \triangle$ are bounded.
	\begin{enumerate}
\item If the perturbation $\triangle$ is such that:\\
	\begin{equation*}
	\triangle =\pi_{\eta_t} \triangle, \mbox{ or equivalently } \eta_t^\top\triangle(t)=0,\ \forall t\ge 0,
	\end{equation*}
then, for any initial condition such that $d_t(0)=-\eta_t^\top \xi_t(0) \in \mathbb{R}^+$,  direct application of the feedback control \eqref{eq:FT} ensures that: i) Item 1 of Theorem \ref{th:landing} holds, ii) $\dot{v}(t)$ and $v(t)$ are bounded and converging asymptotically to zero,  and finally iii) $\| \pi_{\eta_t}\xi_t\|$ is ultimately bounded by $\Delta_\xi$, solution of $\|\pi_{\eta_t} q_t\| = \frac{\|\triangle\|_{\max}}{k^t_{d_{1,2}}}$.
\item If $\eta_t^\top\triangle(t)\neq 0$,  then, for any initial condition such that $d_t(0)=-\eta_t^\top \xi_t(0) \in \mathbb{R}^+$, the following slightly modified feedback control:
\begin{align}
F_t=K^t_pq_t+K^t_d(\phi_t-\eta_t \phi_t^*)+mge_3 \label{eq:FT2}
 \end{align}
 with $\phi^*_t \geq \frac 1 {k_{d_3}^t}|\eta_t^\top\triangle(t)|_{\max}$,
ensures that the above i) and ii) assertions hold and guarantees that $\xi_t$ is bounded.
\end{enumerate} 
\end{prop}
\begin{proof}
	See Appendix B.
\end{proof}
\begin{remark}
The focus of the above proposition is on robustness and adaptation of the controller with respect to the bounded perturbation $\triangle$. It is introduced particularly to show robustness of the proposed control law with respect to bounded perturbations in the plane orthogonal to $\eta_t$ and, in the interest of a less complicated presentation, a slightly modified version of the control law \eqref{eq:FT} is introduced in \eqref{eq:FT2} to be able to analyse  the robustness of the closed loop system with respect to any bounded disturbance.
\end{remark}
\subsection{Going through the center of the window} \label{sec:Fw}
To accomplish the goal of going through the window, while avoiding the wall and window edges, the following control law is proposed
\begin{equation}\small
F_w= \sigma(q_w)(k^w_p\pi_{\eta_w}\bar{q}_w+k^w_d\pi_{\eta_w}\phi_{w}+k^w_\phi\eta_w(\eta{_w}^\top\phi_{w}-\phi^*_w)+mge_3),\label{eq:Fw}
\end{equation}
with $k^w_p$, $k^w_d$ and $k^w_{\phi}$ positive gains, $\phi^*_w>0$ and
\begin{equation}\label{eq:sigma}
\sigma(q_w)=
\begin{cases}
0 &\text{, if } \eta_w^\top q_w \ge 0\\
1&\text{, if } \eta_w^\top q_w < 0,
\end{cases}
\end{equation}
which indicates that when the vehicle already crossed the window ($d_o\le 0$), $F_w=0$.
Note that when $\eta_w^\top q_w < 0$, the resulting closed-loop system can be written as
\begin{equation}\scalebox{0.9}{$
	\left \{
	\begin{aligned}
	\dot\xi_w&=v\\
	\dot v&= -k^w_p\pi_{\eta_w} \frac{\xi_w}{d_w} - k^w_d\pi_{\eta_w} \frac{v}{d_w} -k^w_\phi\eta_w(\eta{_w}^\top \frac{v}{d_w}  -\phi^*_w)+\triangle,
	\end{aligned}\label{eq:closedloop_w}
	\right.$}
\end{equation}
The unknown term $d_w$ is a convex combination of the unknown distances $d_o$ and $d_e$:
\begin{equation}
\frac{1}{d_w} = ( \alpha_w\frac {1} {d_o} +  (1-\alpha_w)\frac {1} {d_e})
\end{equation}
which is deduced from \eqref{d_w} and \eqref{eq:phi_w} according to the definition of $\alpha_w$ \eqref{eq:alphaw}:
\begin{equation}
d_w = \begin{cases}
d_e, &\text{if } \|q_w\| \le \epsilon \ (\xi_w\in \mathcal W) \\
\frac{d_o d_e}{\alpha_w d_e + (1-\alpha_w) d_o}, &\text{if } \epsilon  < \|q_w\| <\epsilon+\delta\\
d_o. &\text{if } \|q_w\| \ge \epsilon +\delta\\
\end{cases}
\end{equation}
\begin{remark}
	Note that the unknown  time varying distance $d_w$ involved in the closed-loop system is due to the use of feedback information $\bar{q}_w=\frac {\xi_w} {d_w}$ and $\phi_w=\frac v {d_w}$  in the control law.  It is the key feature to achieve the double goal of avoiding collision with the wall and window edges as well, while ensuring the main task of going through the center of the window. When
	the vehicle approaches the wall or window edges outside the
	region $\mathcal W$, $d_w = d_o$. If $d_o$ is decreasing then the motion in the orthogonal direction to the wall is highly damped while the region $\mathcal W$ is highly attractive. In practice, this leads to a bounded high
	gain in the feedback control that prevents collision. When the vehicle is inside the region $\mathcal W$, $d_w=d_e$. This later is lower bounded by a positive constant so that the vehicle is able to go through the center of the window with a non-zero velocity. More details of analysis will be shown below.
\end{remark}
\begin{prop} \label{prop:window}
	Consider the system (\ref{eq:system}) with the control input given by \eqref{eq:Fw}.
	If the positive gains $k^w_p$, $k^w_d$ and $k^w_\phi$ are such that $\frac{{k_d^w}^2}{k_p^w}>\frac {r_w} 2$ and for any arbitrary small $\epsilon>0$, the chosen $\phi^*_w$ satisfies:
	\begin{equation}
	\phi^*_w>\frac{|\eta_w^\top\triangle(t)|_{\max}}{k^w_\phi}+\epsilon, \ \forall t\ge 0,
	\end{equation}
	then for any initial condition satisfying $d_w(0)\in \mathbb R^+$ and as long as $\sigma(q_w(t))=1$, the following assertions hold $\forall t\ge 0$:
	\begin{enumerate}
		\item there exists a finite time $t_w\geq0$ at which the vehicle enters the region $\mathcal W$ ($\|q_w(t_w)\|\le \epsilon$) and remains there, while $d_w(t)\ge d_o(t)\in \mathbb R^+,\ \forall t<t_w$,
		\item there exists a finite time $t_{\lim}> t_w$ at which the vehicle crosses the window $d_o(t_{\lim}) = 0$, with strictly negative velocity $\dot d_o(t_{\lim})$ such that the vehicle is inside the region $\mathcal W$ ($\|q_w(t)\| \le \epsilon$) for all $t \in [t_w, t_{\lim})$.
	\end{enumerate}
\end{prop}
\begin{proof}
	See Appendix C.
\end{proof}

\subsection{Application Scenario}\label{subsec:application}

The double goal of crossing the window and landing on the landing pad can be achieved by simply applying the control laws $F_w$ and $F_t$ in sequence, with an adequate trigger to switch from $F_w$ to $F_t$. Taking the limitation of the cameras' field of view into the consideration, there will be four different modes during the full process of going through a window and landing on the pad. When $t\in[T_1,T_2)$, $mode=1$ and $F_w$ \eqref{eq:Fw} is active. When the vehicle approaches to the center of the window, the on-board camera loses the full image of the window and $mode$ changes to 2. When $t\in[T_2,T_3)$, $mode$ $2$ is active and the open-loop control $\eta_w|\eta_w^\top F_w(T_2^-)|$ is applied, where  $T_2^-$ is the last time instance before the camera loses the image of the window. At the time instance $t=T_3$, when the downward-looking camera detects the landing pad, the $mode$ changes to 3 and the control law $F_t$ \eqref{eq:FT} is applied when $t\in[T_3,T_4)$. At time instance $t=T_4$, the vehicle is already close to the center of the landing target and it is safe to slowly shutdown the quadrotor motors. In order to avoid inadequate behaviors, the switch from $mode$ 2 to 3 is only triggered once. Moreover, in practice, due to the limitation of camera's field of view, the initial errors should not be large and should converge to zero fast enough, thereby allowing the vehicle to almost align with the center of the window before switching to $mode$ 2. 
Additionally, the position of the landing target should be close enough to the window so that the quadrotor is able to timely detect the landing target after it goes through the window. The switching between different modes is based on the combination of selected frames from both the downward-looking and forward-looking on-board cameras obtained in the experiments. The detail on the adopted procedure is described in Section \ref{sec:experiment}.

\section{Simulation Results}\label{sec:simu}
In this section, simulation results are presented to illustrate the behavior of the closed-loop system using the proposed controller. A high-gain inner-loop controller is used to control the attitude dynamics (\cite{tang2015homing}). It generates the torque inputs in order to stabilize the orientation of the vehicle to a desired one defined by the desired thrust direction $R_de_3$, which is provided by the outer-loop image-based controller, and the desired yaw chosen to align the forward-looking camera with direction orthogonal to the wall.
 The control algorithm is tested with different initial conditions, always starting from a position outside the room containing the target (see Fig. \ref{fig:3D}). The initial velocity of the quadrotor is $v(0)=[0\ 0\ 0]^\top$, and the gains are chosen as $K_p^t=\text{diag}[4\ 4\ 1.75]$, $K_d^t=4 I_3$, $k_d^w=0.8$, $k_p^w=1$, $k_\phi^w=1$ and $\phi^*_w=0.3$

\begin{figure}[!htb]
	\centering
	\includegraphics[scale = 0.65]{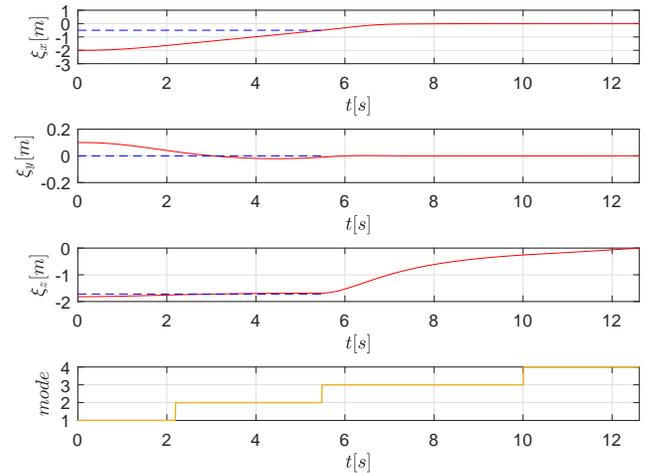}
	\caption{Evolutions of the quadrotor's position $\xi$ and the mode}
	\label{fig:P}
\end{figure}

\begin{figure}[!htb]
	\centering
	\includegraphics[scale = 0.65]{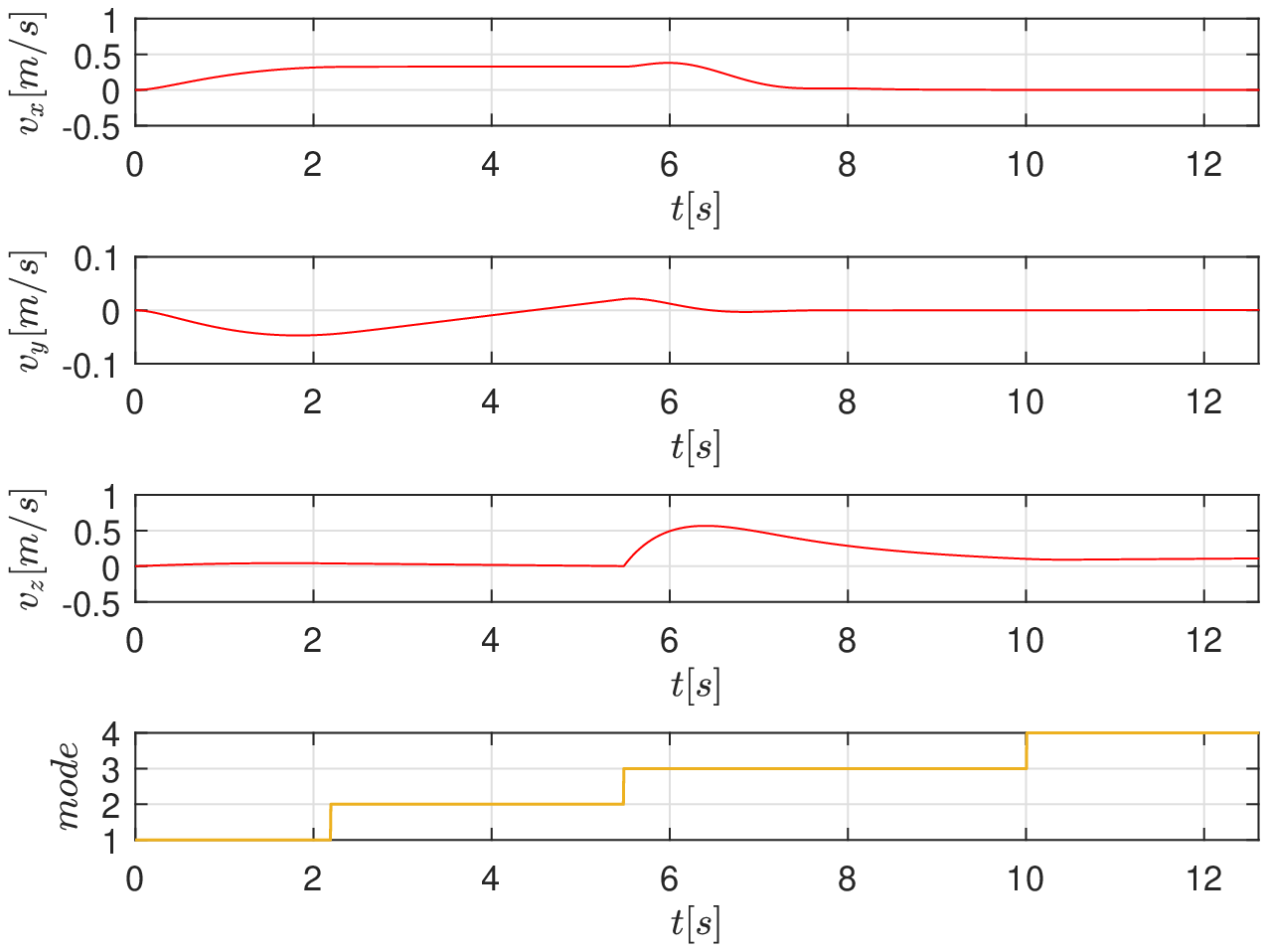}
	\caption{Evolutions of the quadrotor's velocity $v$ }
	\label{fig:V}
\end{figure}

\begin{figure}[!htb]
	\centering
	\includegraphics[scale = 0.65]{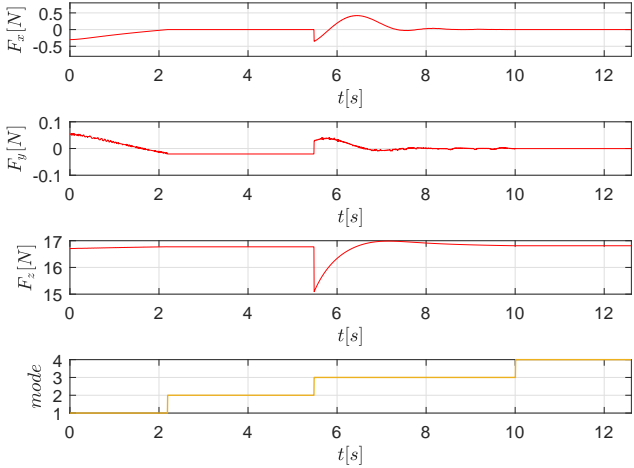}
	\caption{Evolutions of the virtual control input $F$ }
	\label{fig:F}
\end{figure}

\begin{figure}[!htb]
	\centering
	\includegraphics[scale = 0.65]{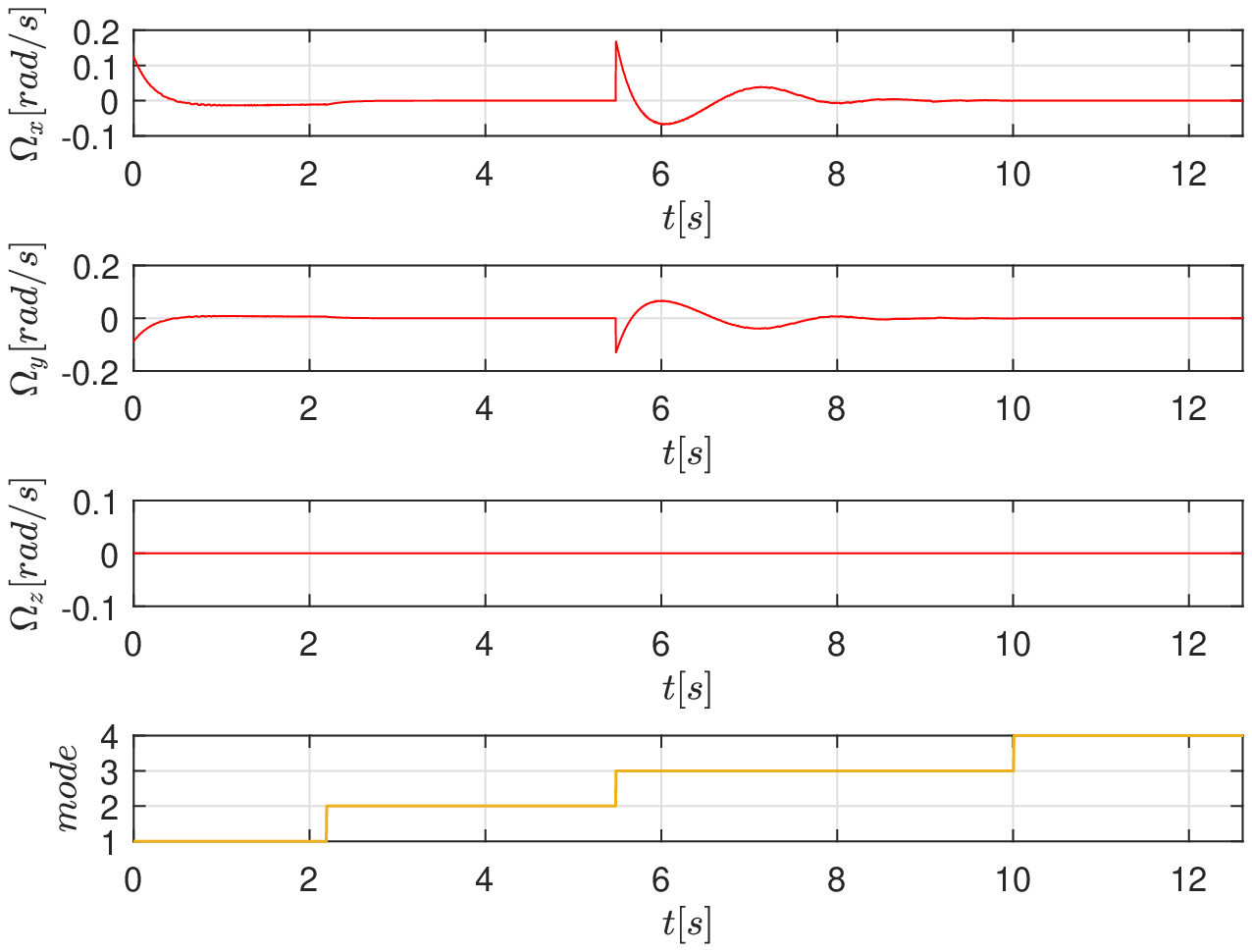}
	\caption{Evolution of angular velocity $\Omega$ }
	\label{fig:omega}
\end{figure}

\begin{figure}[!htb]
	\centering
	\includegraphics[scale = 0.65]{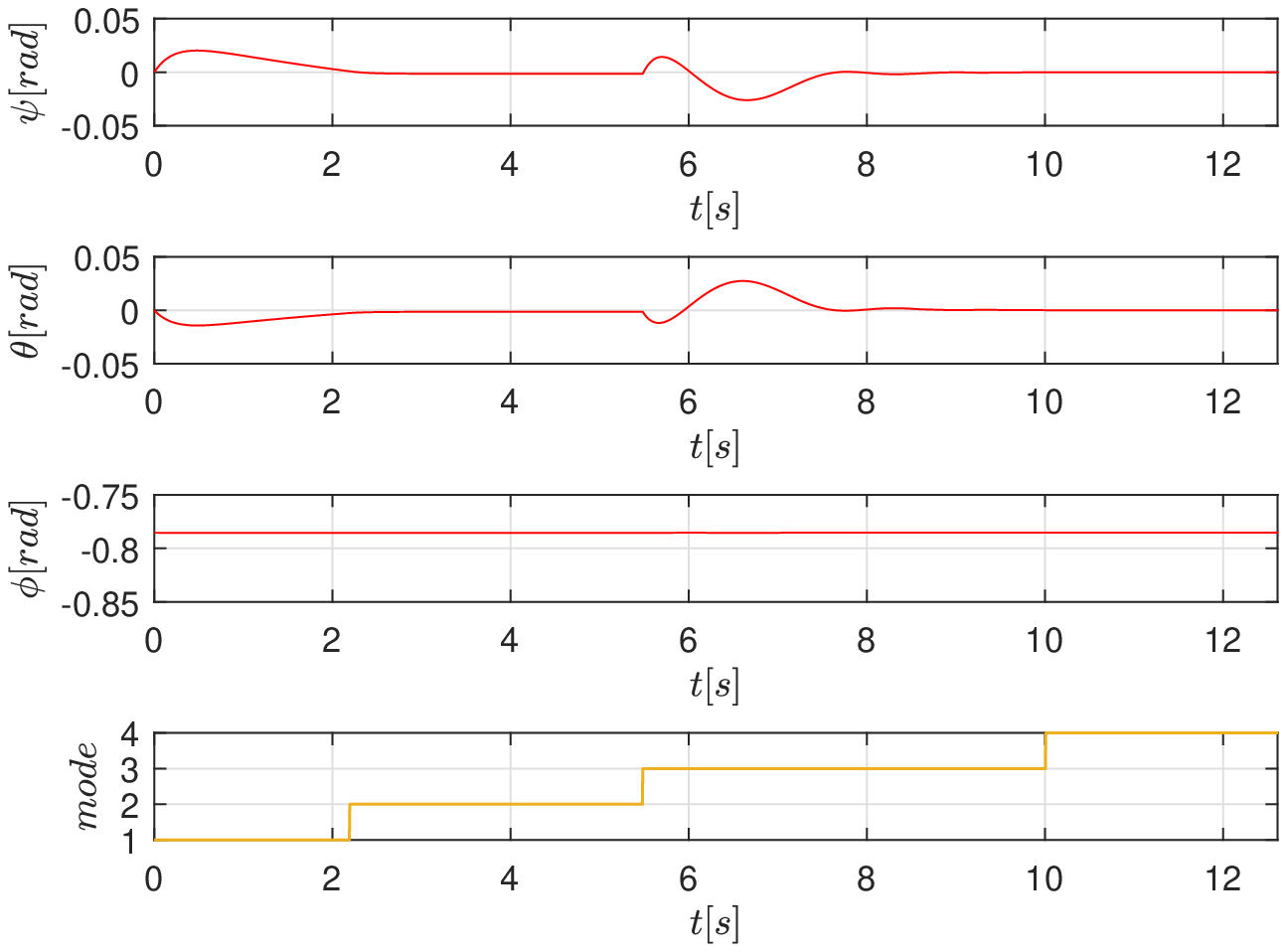}
	\caption{Evolution of Euler angles }
	\label{fig:eular}
\end{figure}

\begin{figure}[!htb]
	\centering
	\includegraphics[scale = 0.65]{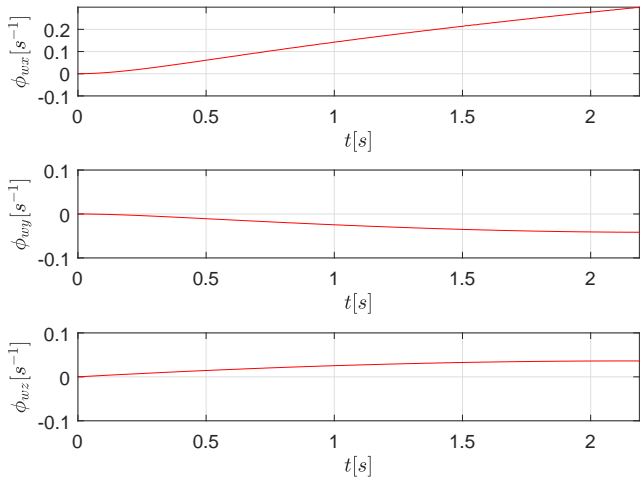}
	\caption{Translational optical flow using for going through the window during mode $1$.}
	\label{fig:OF_w}
\end{figure}

\begin{figure}[!htb]
	\centering
	\includegraphics[scale = 0.65]{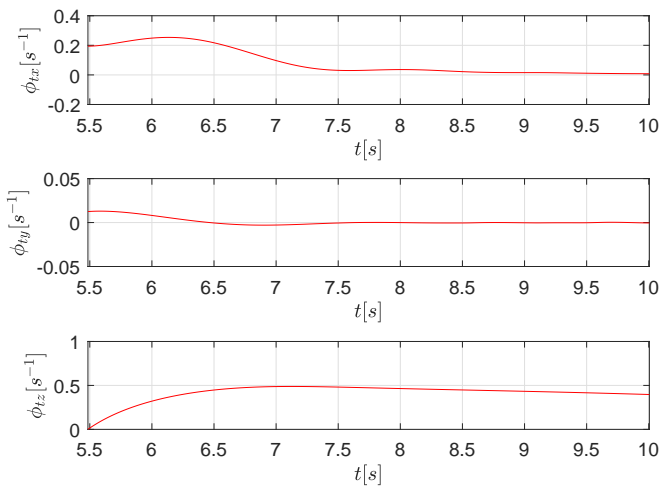}
	\caption{Translational optical flow using for landing during mode $3$.}
	\label{fig:OF_T}
\end{figure}

\begin{figure}[!htb]
	\centering
	\includegraphics[scale = 0.65]{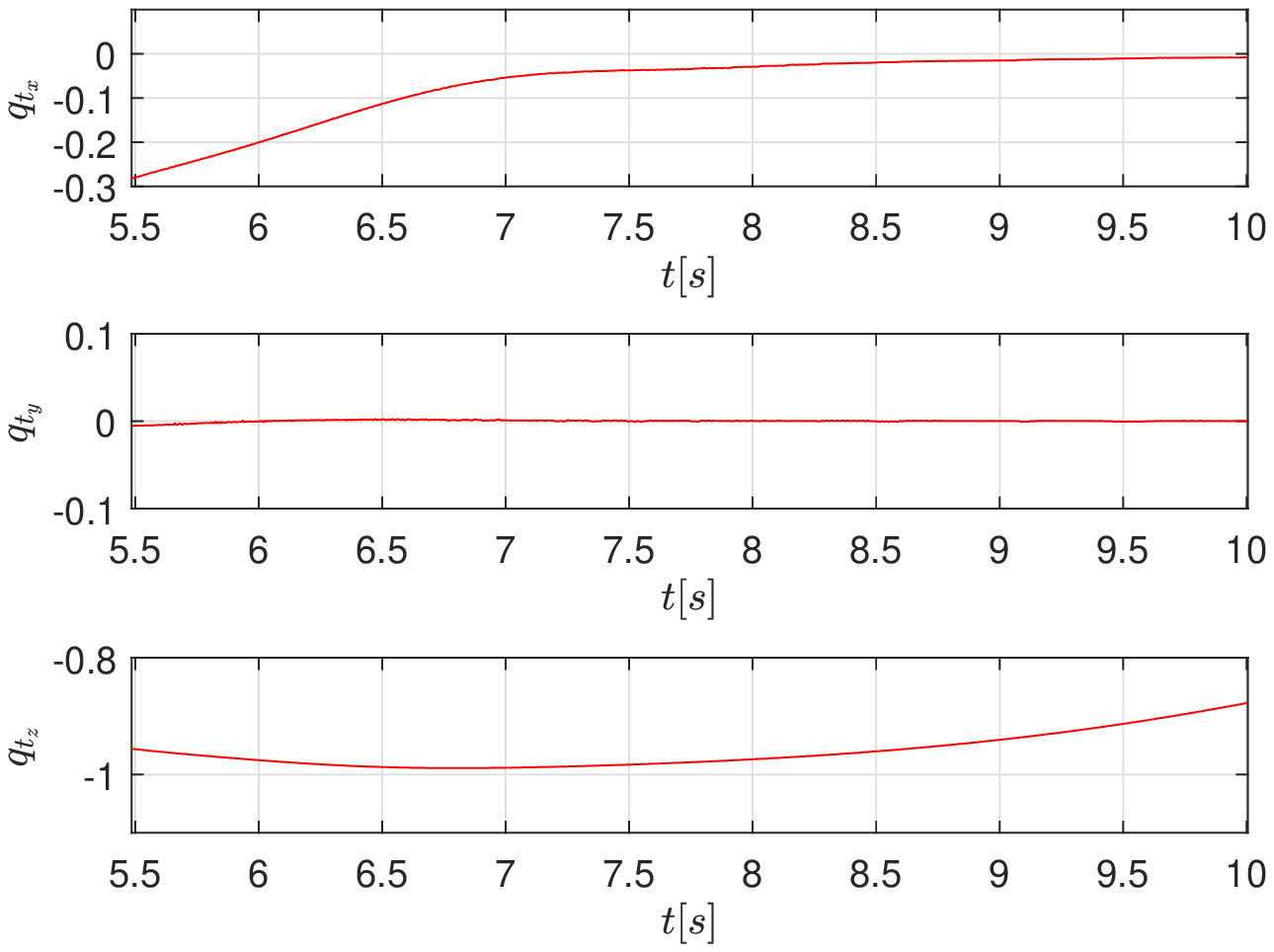}
	\caption{Image feature $q_t$ during mode $3$.}
	\label{fig:q_T}
\end{figure}

As shown in Fig. \ref{fig:3D}, with different initial positions the quadrotor successfully avoids the wall and window,  goes through the center of the window, and then lands on the center of the landing target. Figures \ref{fig:P}-\ref{fig:q_T} show in detail the time evolution of quadrotor's state variables, virtual input, and image features for the initial position $\xi(0)= [-2\ 0.1\ -1.82]^\top$. The time evolution of the active mode is also specified. In mode 1, the quadrotor is approaching the window; in mode 2, it is crossing the window with no image cues; in mode 3, it starts detecting the landing pad and transitions to the landing maneuver; and finally in mode 4, the motors are shutdown.

Figure \ref{fig:P} shows the time evolution of the vehicle's position and the dashed lines are the coordinates of window's center. From Figure \ref{fig:P}, one can see the quadrotor first converges to the center line of the window and then converges to the center point of the target. Fig. \ref{fig:V} shows the time evolution of the vehicle's velocity. The virtual control input $F$ is shown in  Fig.~\ref{fig:F}. The angular velocity of the quadrotor is depicted in Fig.~\ref{fig:omega} and Fig. \ref{fig:eular} depicts the time evolution of Euler angles, which indicates a good compromise in terms of time-scale separation between the outer-loop and inner-loop controller. Figures \ref{fig:OF_w} and \ref{fig:OF_T} show the translational optical flow used for going through the window in mode 1 and for landing in mode 3, respectively. The evolution of image features of $\bar{q}_w$ and $q_t$ are depicted in Figures \ref{fig:bar{q}_w} and \ref{fig:q_T}, respectively. We can see that the image features $\bar{q}_w$ and $q_t$ approach to the desired values $[-1\ 0\ 0]^\top$ and $[0\ 0 \ 0]^\top$, respectively, before the on-board cameras lose the image information.
\begin{figure}[!htb]
	\centering
	\includegraphics[scale = 0.65]{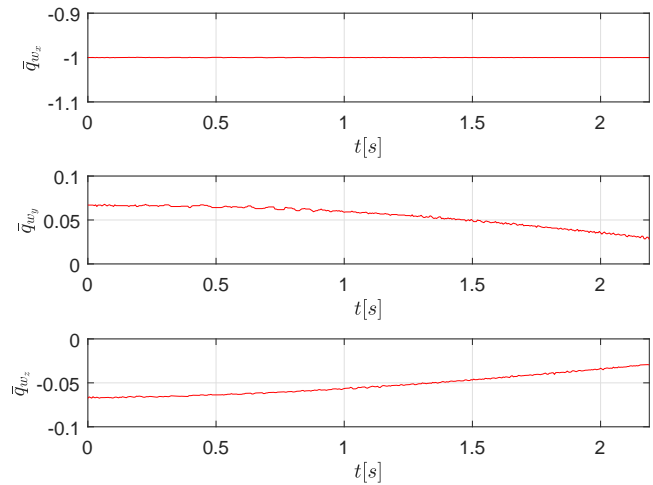}
	\caption{Image feature $\bar{q}_w$ during mode $1$.}
	\label{fig:bar{q}_w}
\end{figure}
\section{Experiments} \label{sec:experiment}
\subsection{Experimental setup}
\begin{figure}[!htb]
	\centering
	\includegraphics[scale = 0.045]{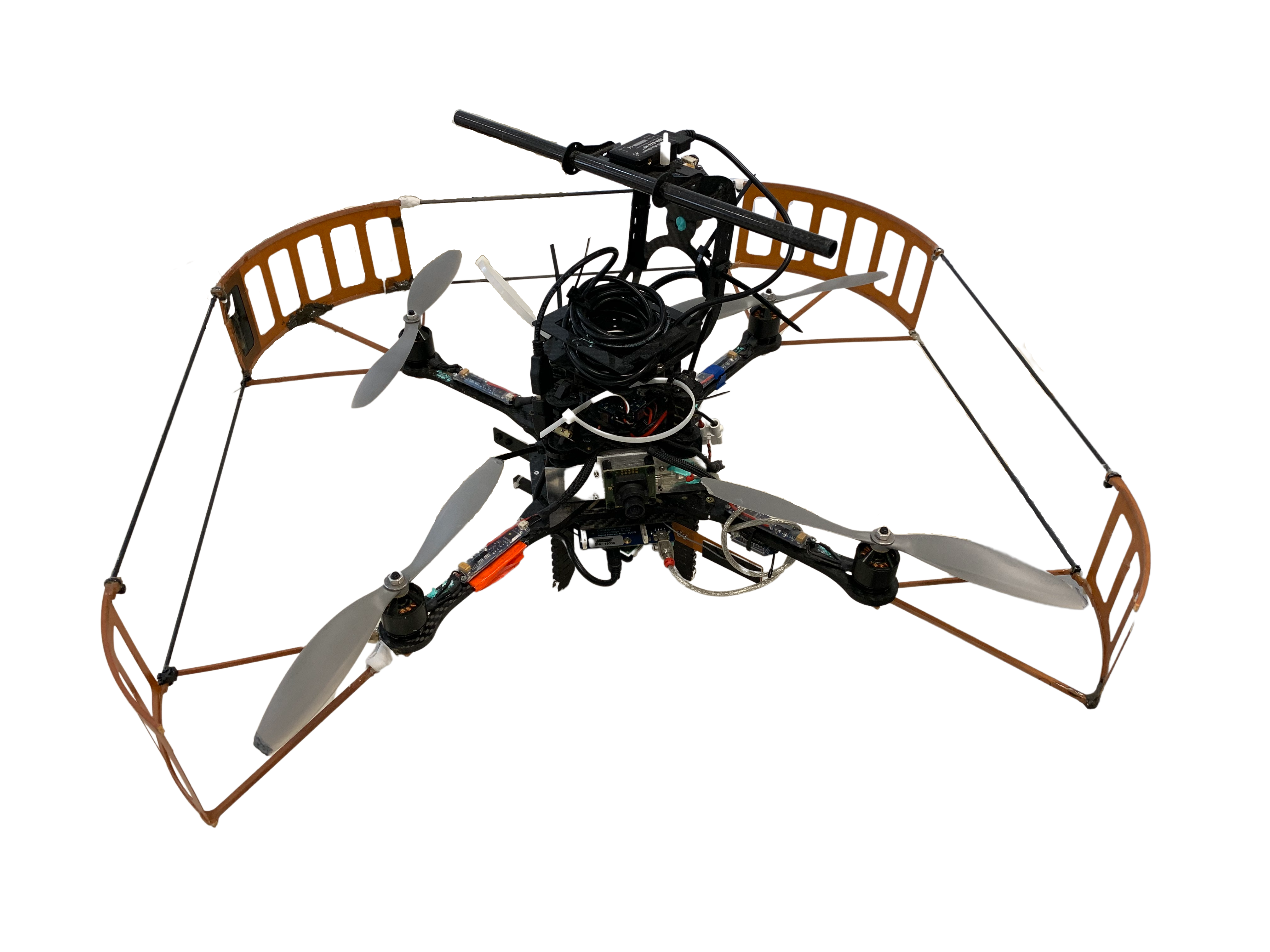}
	\caption{Asctec Pelican quadrotor.}
	\label{fig:astec}
\end{figure}

\begin{figure}[!htb]
	\centering
	\includegraphics[scale = 0.26]{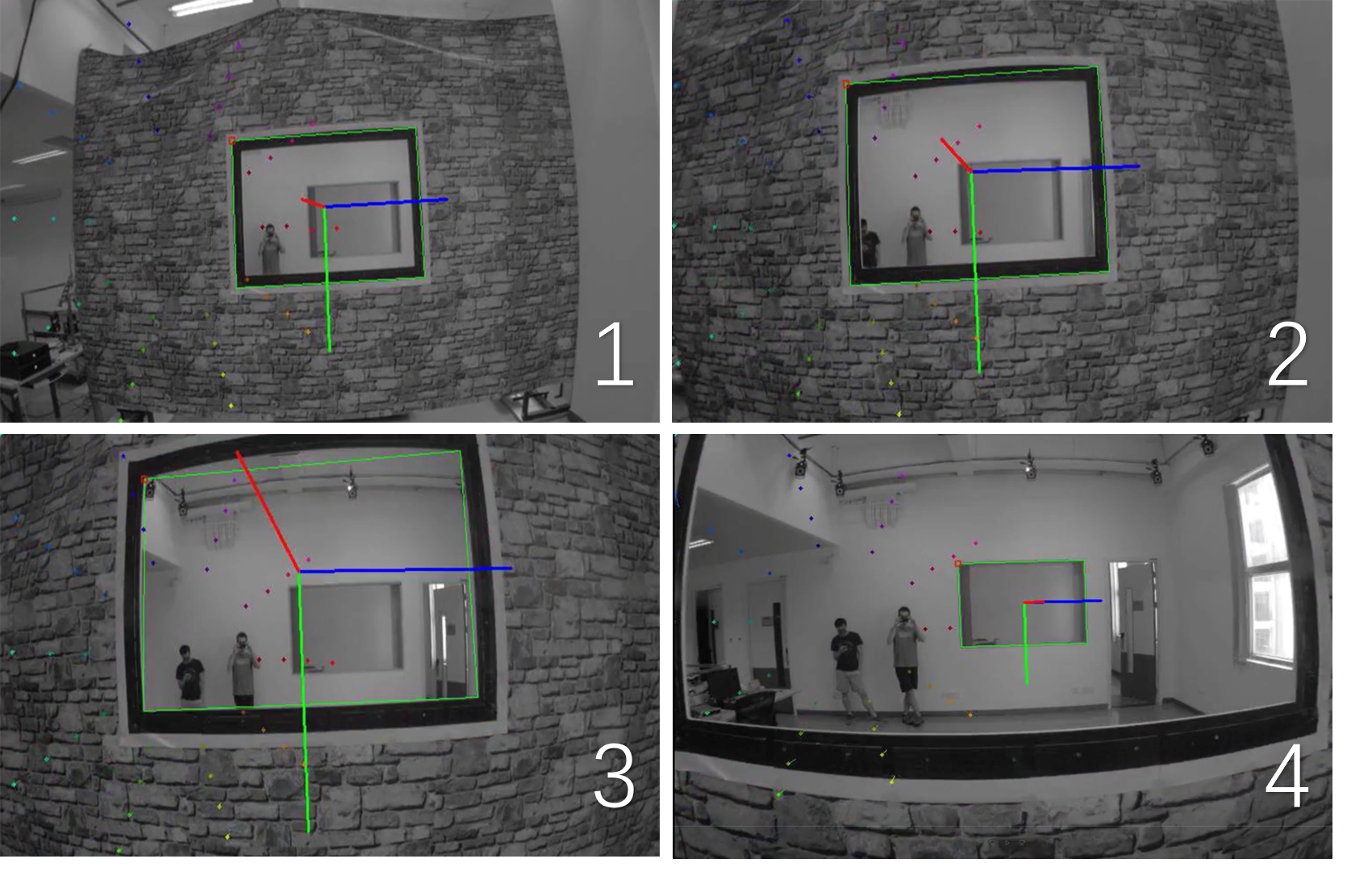}
	\caption{Selected frames from the forward-looking camera.}
	\label{fig:camera}
\end{figure}

In order to set up the experiment, a movable wall was used to divide the testing space into two smaller compartments and a landing pad was placed on the ground of the second one. The partition wall contains a rectangular window and is textured as a brick wall to provide the background optical flow, as shown in Fig.~\ref{fig:camera}.
The vehicle used for the experiments is an Asctec Pelican quadrotor (Fig. \ref{fig:astec}) with weight 1676g and the arm length from the center of mass to each motor is 20cm. The available commands are thrust force and attitude which are derived from the force $F$ provided by the outer-loop controller \eqref{eq:FT} (respectively \eqref{eq:Fw}) and the desired yaw angle. The quadrotor is equipped with two wide-angle cameras, one pointing towards the ground and another is facing the forward direction, pointing at the wall. Recalling Assumption \ref{assum:1}, the downward-looking camera reference frame coincides with the vehicle's body fixed frame and  the rotation matrix from the forward-looking camera reference frame to the body frame is ${_{C}^{B}}R=R_Z(-\frac{\pi}{4})R_X(\frac{\pi}{2})$. These two cameras are uEye UI-122ILE models featuring a 1/2-in sensor with global shutter which operate at a resolution of $752\times 480$ pixel at 50 frames per second and are provisioned with 2.2-mm lenses.

In the experiments, a rapid prototyping and testing architecture are used in which a MATLAB/Simulink environment integrates the sensors and the cameras, the control algorithm and the communication with the vehicle. The controller is developed and tuned on a MATLAB/Simulink environment and C code
is generated and compiled to run onboard the vehicle as a final step. 
The onboard computer (a 4-core Intel i7-3612QE at 2.1GHz, named AscTec Mastermind) is responsible for running in Linux three major software components that provide:
\begin{enumerate}
	\item interface with the camera hardware, image acquisition, feature detection and optical flow computation;
	\item computation of the vehicle force references from the image features, translational optical flow, and angular velocity and rotation matrix estimates provided by the IMU;
	\item interface with microprocessor, receiving IMU data and sending force references to the inner-loop controller.
\end{enumerate}

\begin{figure}[!htb]
	\centering
	\includegraphics[scale = 0.14]{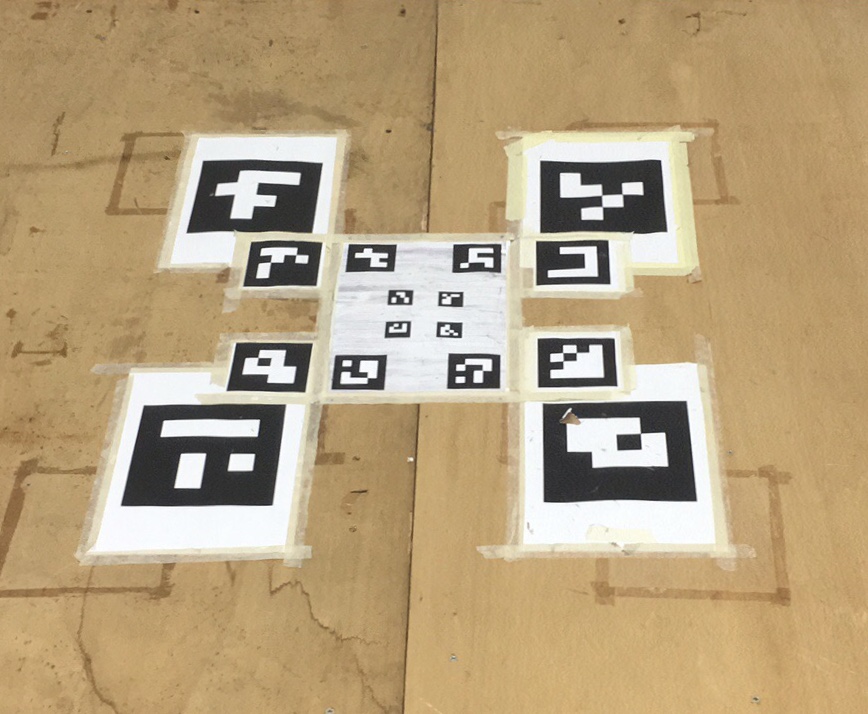}
	\caption{ARUCO markers on the landing pad.}
	\label{fig:landmark}
\end{figure}

\begin{figure}[!htb]
	\centering
	\includegraphics[scale = 0.26]{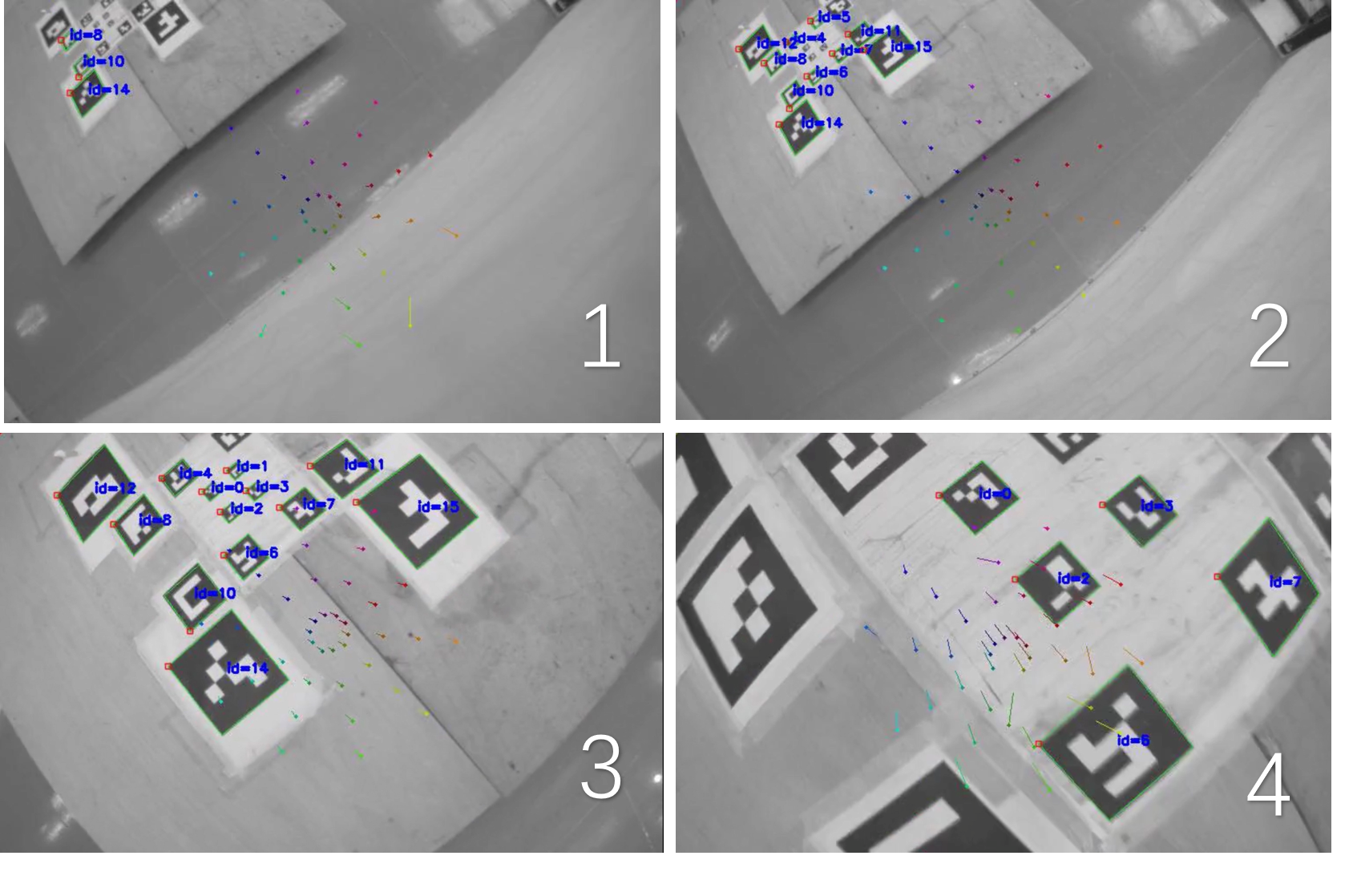}
	\caption{Selected frames from the downward-looking camera.}
	\label{fig:camera2}
\end{figure}

A Python program running on the onboard computer performs detection of the window, detection of landing target, and optical flow computation using the OpenCV library.
ARUCO markers, for which built-in detection functions exist in the OpenCV library, are used to define the landmarks on the landing pad. In order to fit the camera's field of view during the full process of landing, the landmarks are composed by 4 groups of ARUCO markers and in each group there are 4 ARUCO markers with same border size but different identifier (id) as shown in Fig.~\ref{fig:landmark}.
When the camera is far away from the markers, the group of larger markers can be seen and when the camera is near the ground, only the smaller group of the landmarks will be shown in the field of view.  The rectangular window shape is detected using the library code originally developed for ARUCO marker's border detection. The detected window frame (in green) and the window's coordinate system overlayed on the image are show in Fig.~\ref{fig:camera} (1), (2), and (3). The translational optical flow is also computed onboard. The computation is based on the conventional image plane optical flow field provided by a pyramidal implementation of the Lucas-Kanade algorithm. The detailed description of the computation can be found in \cite{serra2016landing}. The small vectors represented in Fig. \ref{fig:camera2} represent the translational optical flow of the image pixels.

In order to provide ground truth measurements and evaluate the performance of the proposed controller, a VICON motion capture system (\cite{vicon}) which comprises 12 cameras is used together with markers attached to the quadrotor, window, and landing target. The motion capture system is able to accurately locate the position of the markers, from which ground truth position and orientation measurements are gathered. Note that, none of the measurements from the motion capture system are used in the proposed controller.
\subsection{Experimental results}
The experiments were conducted with the same control gains as the simulations. Before the proposed controller is triggered, the vehicle is hovering at position $\xi=[0.15,1.79,-1.76]$~m, which is outside the space containing the landing pad. As mentioned in Section \ref{subsec:application}, there are four different modes during the full process of going through a window and landing on the target due to the limitation of the field of view of the on-board cameras. Fig.~\ref{fig:camera} shows the selected frames in a timed sequence from the forward-looking camera. These four frames are a fixed time step apart and were taken during a $mode$ 1 to $mode$ 2 transition. In Fig.~\ref{fig:camera}~(1), (2), and (3), $mode$ 1 is active, and one can see that the window frame is well detected. In Fig.~\ref{fig:camera}~(1), $t=T_1$ and the controller $F_w$ is triggered. In Fig.~\ref{fig:camera}~(2) and (3), the vehicle is still in $mode$ 1 and approaches the center of the window. As the vehicle approaches the window, the window frame disappears from the field of view of the camera and at time $t=T_2$ the $mode$ commutes to 2, as shown in Fig.~\ref{fig:camera}~(4). Note that during transition from $mode$ 1 to 2, instead of losing the window frame, the camera may detect rectangles other then the target window, as depicted in Fig.~\ref{fig:camera}~(4). In order to avoid this situation, the $mode$ changes from 1 to 2 if the pixel coordinates change instantaneously in a way that is incompatible with smooth tracking of the same window object. Fig.~\ref{fig:camera2} shows the selected frames in a timed sequence from the downward-looking camera. These four frames were taken at fixed time steps during a transition from $mode$ 2 to 3 to 4. As shown in Fig.~\ref{fig:camera2}~(1), the vehicle has already crossed the window but the landing pad is not fully detected thus the $mode$ is still 2. At time instance $t=T_3$, as shown in Fig.~\ref{fig:camera2}~(2), the downward-looking camera detects successfully the landing pad, the $mode$ is switched to 3 and $F_t$ is applied as control input. Recall that the switching from $mode$ 2 to $mode$ 3 is only triggered once in order to avoid inadequate behavior. In Fig.~\ref{fig:camera2}~(3), the vehicle approaches the target and the $mode$ is still $3$. At the time instance $t=T_4$, when the quadrotor has almost reached the target position (see Fig.~\ref{fig:camera2}~(4)), the $mode$ changes to 4 and it is safe to slowly shutdown the motors.    

\begin{figure}[!htb]
	\centering
	\includegraphics[scale = 0.65]{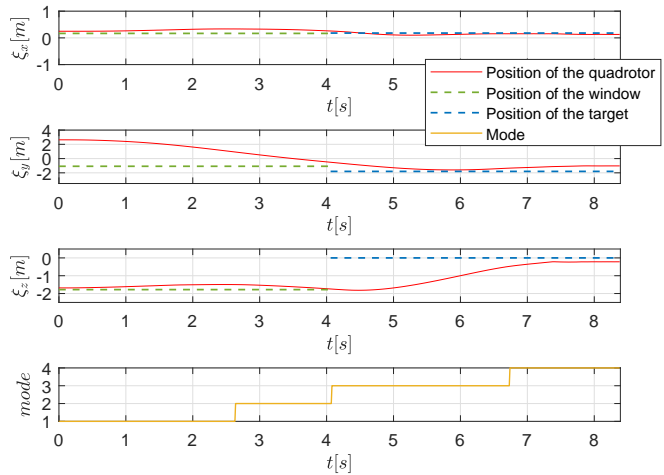}
	\caption{Evolutions of the quadrotor's position $\xi$ and the mode}
	\label{fig:P_exp}
\end{figure}

\begin{figure}[!htb]
	\centering
	\includegraphics[scale = 0.65]{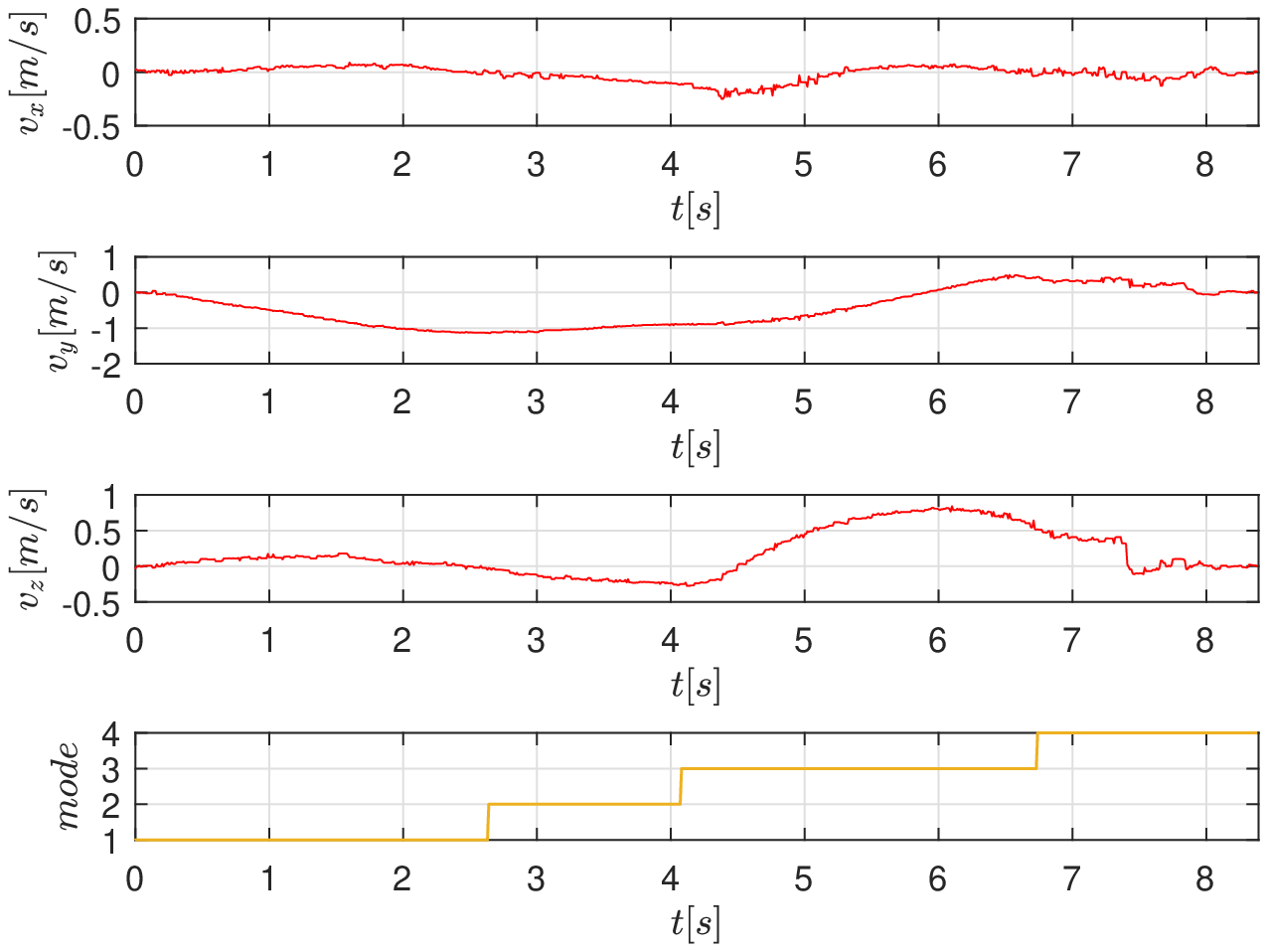}
	\caption{Evolutions of the quadrotor's velocity $v$ }
	\label{fig:V_exp}
\end{figure}

\begin{figure}[!htb]
	\centering
	\includegraphics[scale = 0.65]{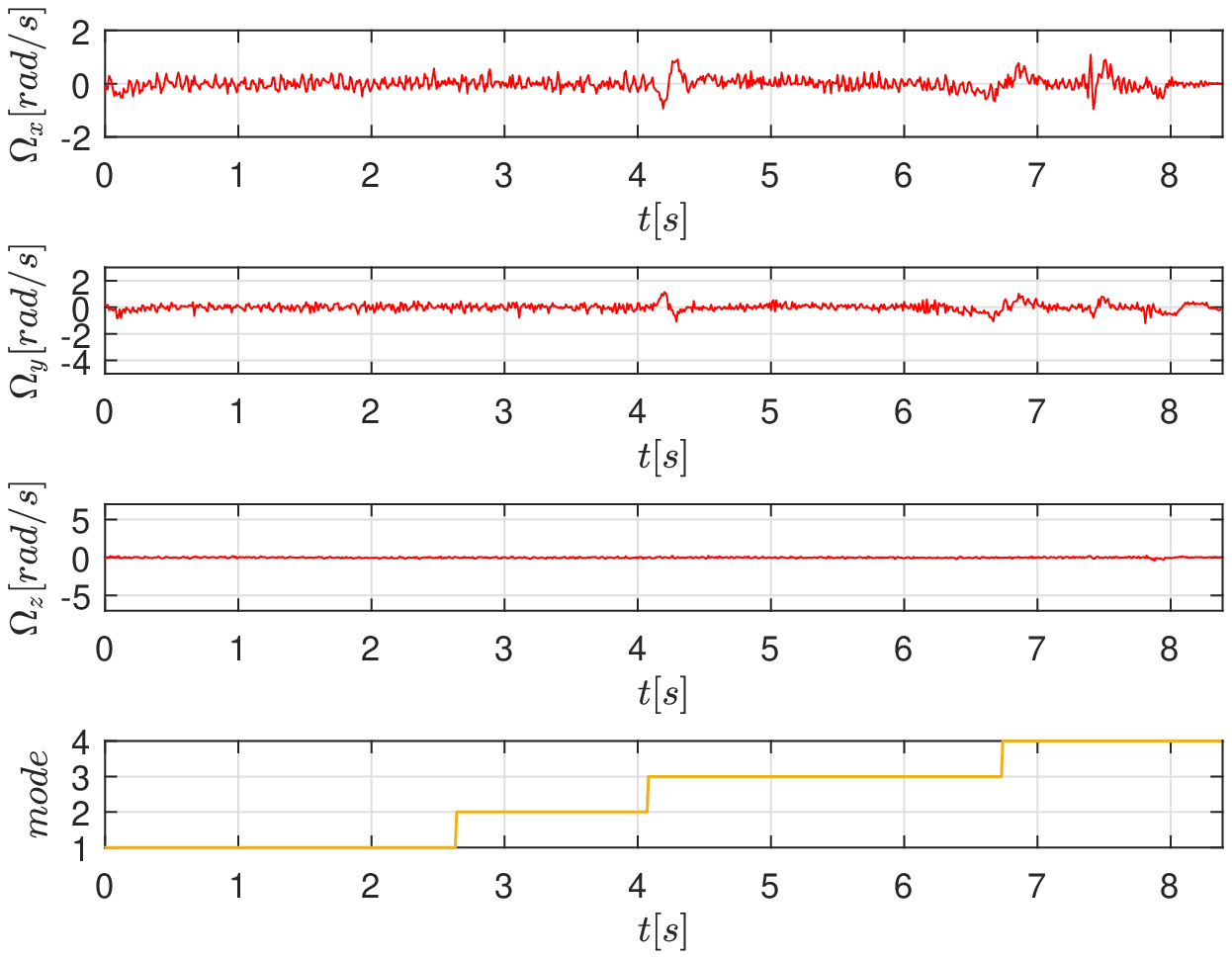}
	\caption{Evolution of angular velocity $\Omega$ }
	\label{fig:omega_exp}
\end{figure}

\begin{figure}[!htb]
	\centering
	\includegraphics[scale = 0.65]{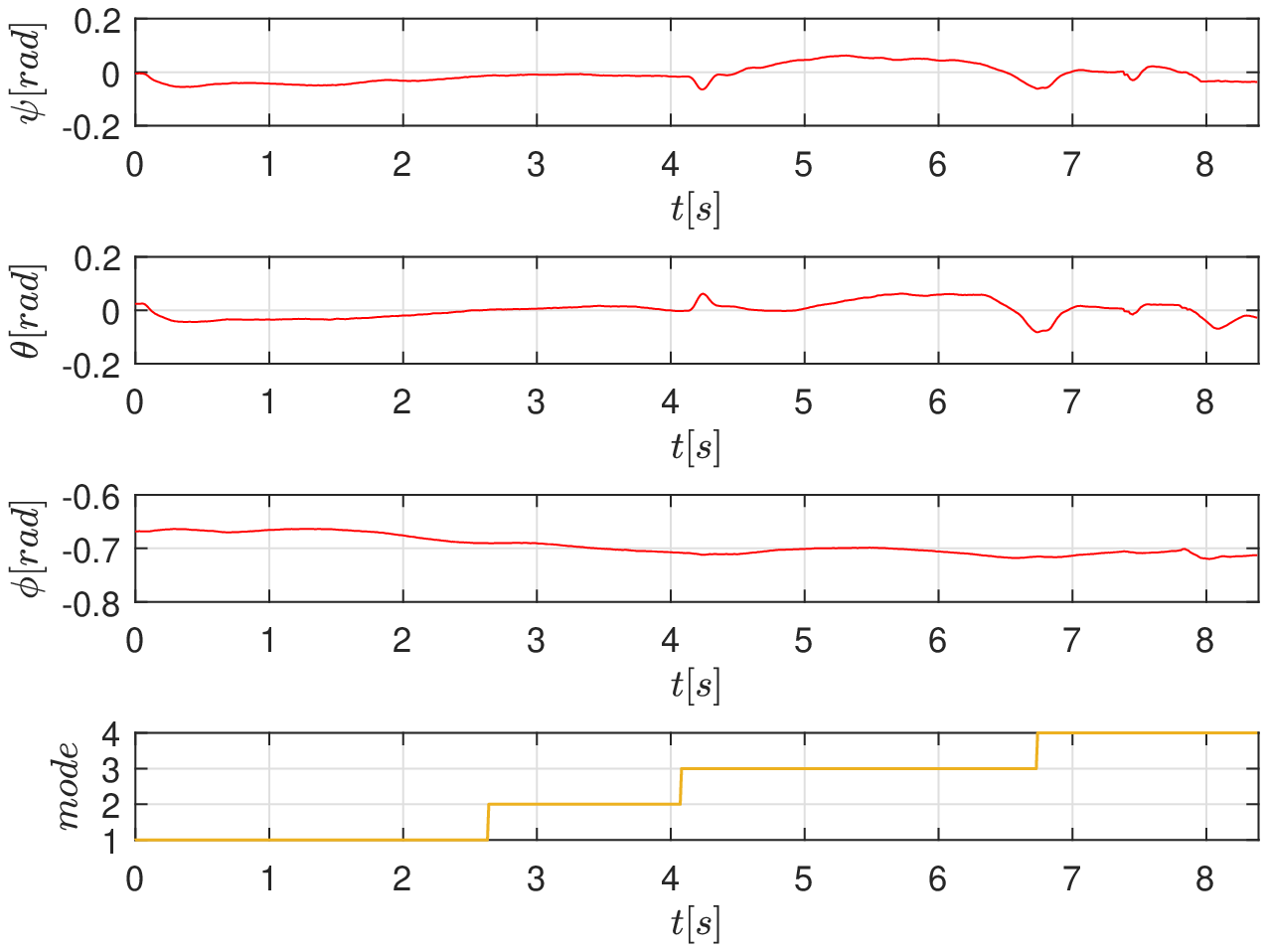}
	\caption{Evolution of euler angle }
	\label{fig:euler_exp}
\end{figure}

\begin{figure}[!htb]
	\centering
	\includegraphics[scale = 0.65]{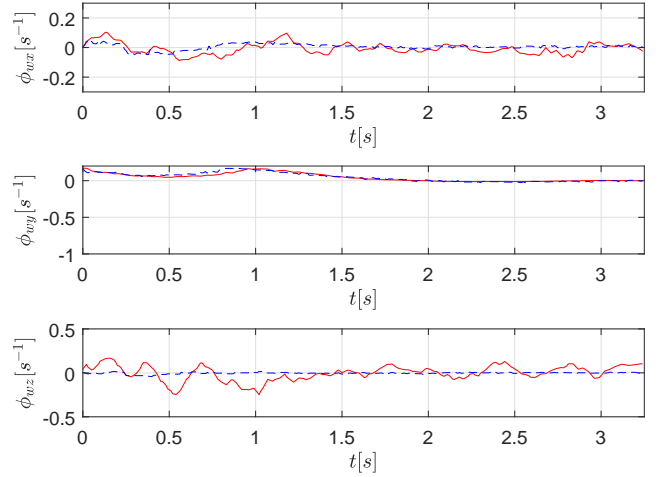}
	\caption{Translational optical flow computed during $mode$ $1$ (going through window) from the image sequence (solid line) and from the VICON measurements (dashed line).}	
	\label{fig:OF_w_exp}
\end{figure}

\begin{figure}[!htb]
	\centering
	\includegraphics[scale = 0.65]{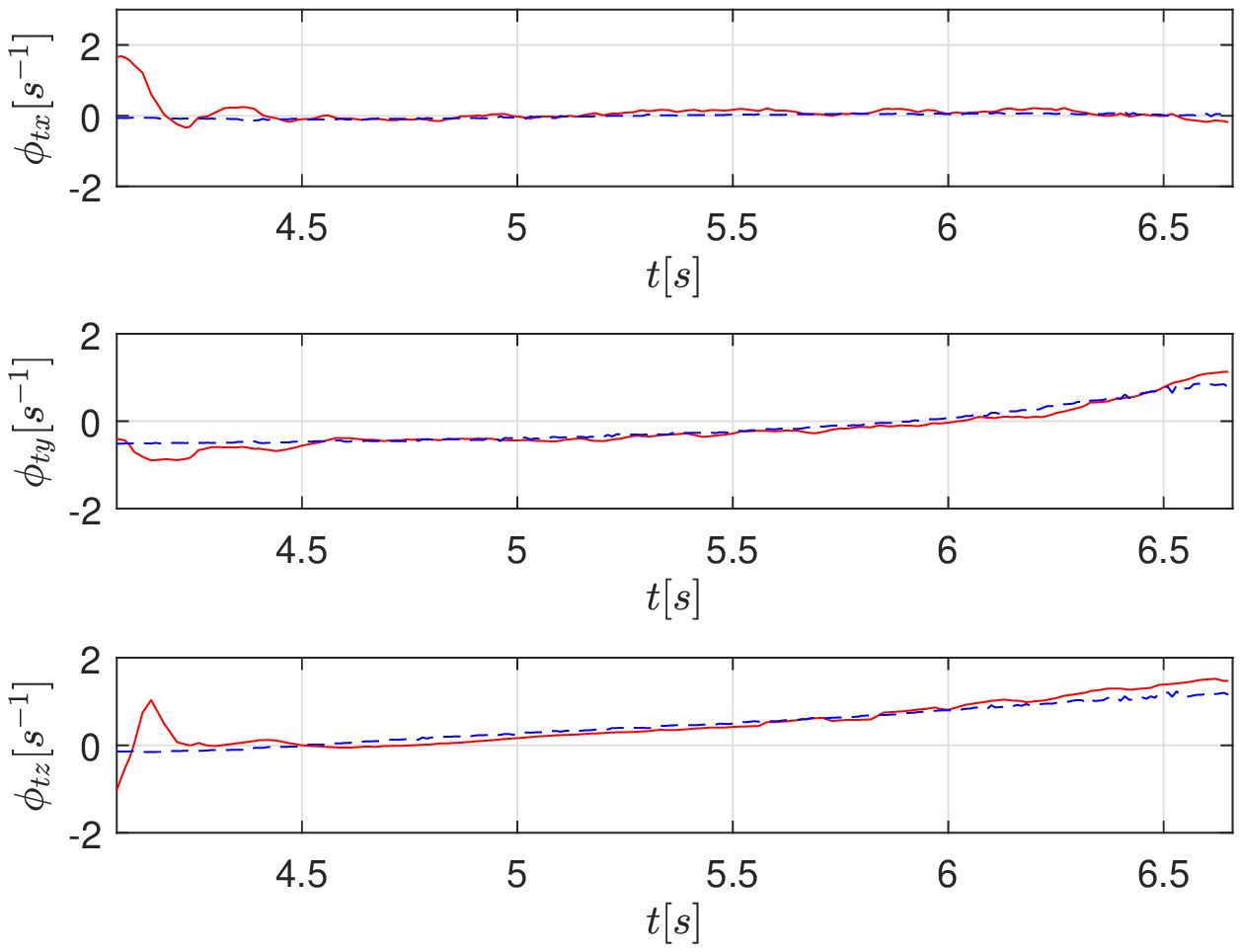}
	\caption{Translational optical flow computed during $mode$ $3$ (landing) from the image sequence (solid line) and from the VICON measurements (dashed line).}
	\label{fig:OF_t_exp}
\end{figure}

Figures \ref{fig:P_exp} and \ref{fig:V_exp} show the position and velocity coordinates of the vehicle provided by VICON, respectively. We can see that the vehicle goes through the center of the window at the end of $mode$ 2 and finally lands on the target. Fig.~\ref{fig:omega_exp} shows the evolution of the angular velocity and Fig.~\ref{fig:euler_exp} show the evolution of the Euler angles. From Fig.~\ref{fig:euler_exp}, one can see that a good compromise in terms of time-scale separation between the outer-loop and
inner-loop controllers is attained, which indicates that the inner-loop controller is sufficiently fast to track the outer-loop references, including during the transitions between different modes. Figure \ref{fig:bar{q}_w_exp} shows the image feature $\bar{q}_w$ used for going through the window. The solid line represents $\bar{q}_w$ computed from the image sequence and the dashed line represents $\bar{q}_w$ provided by the VICON system. There are slight differences between these two computations due to the fact that rotation matrix $R$ provided by the IMU is affected by the surrounding magnetic field generated by the fast rotating motors. Figures \ref{fig:OF_t_exp} and \ref{fig:OF_w_exp} show the translational optical flow used for going through the window and for landing respectively. The solid red lines represent the translational optical flow computed from the image sequence and the dashed line represents the translational optical flow derived from VICON measurements. The video of the experimental results can be found in \url{https://youtu.be/DbpeGfJMHk0}.
\begin{figure}[!htb]
	\centering
	\includegraphics[scale = 0.65]{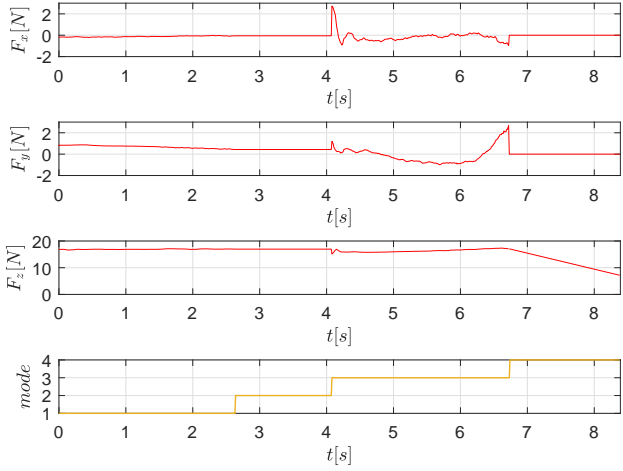}
	\caption{Evolutions of the controller output $F$ }
	\label{fig:F_exp}
\end{figure}
\begin{figure}[!htb]
	\centering
	\includegraphics[scale = 0.65]{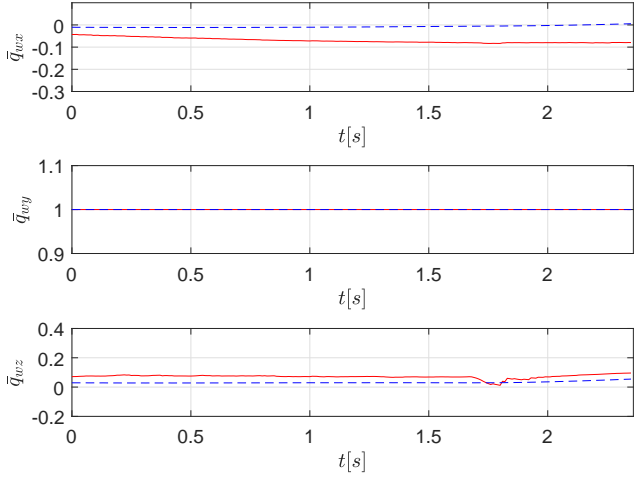}
	\caption{Image features $\bar{q}_w$ computed from the image sequence (solid line) and from the VICON measurements (dashed line) during mode $1$.}
	\label{fig:bar{q}_w_exp}
\end{figure}

\section{Conclusion}\label{sec:conclusion}
This paper considers the problem of controlling a quadrotor to go through a window and land on planar target, using an Image-Based Visual Servo (IBVS) controller. For control purposes,  the centroids vectors provided by the combination of the corresponding spherical image measurements of landmarks (corners) for both the window and the target are used as position feedback. The translational optical flow relative to the wall, window edges, and landing plane is used as velocity measurement. To achieve the proposed objective, no direct measurements of position or velocity are used and no explicit estimate of the height above the landing plane or of the distance to the wall is required. With the initial position outside the room containing the target, the proposed control law guarantees that the quadrotor aligns itself with the center line orthogonal to the window, crosses it with non-zero velocity and finally lands on the planar target successfully without colliding the wall or the edges of the window. The proof of convergence of the overall control scheme is provided and the simulation and experimental results show the effectiveness of the proposed controller.

\appendix
\subsection{Proof of Theorem \ref{th:landing}}
\begin{proof}
	The proof follows a reasoning very similar to that of Theorem 1 in \cite{rosa2014optical}.
	Recalling \eqref{eq:system} and applying the control input \eqref{eq:FT}, the closed-loop system can be written as
	\begin{equation}
	\left \{
	\begin{aligned}
	\dot\xi_t&=v\\
	\dot v&=-K^t_pq_t(\xi_t)-K^t_d\frac{v}{d_t}.
	\end{aligned}
	\right. \label{eq:sys-t}
	\end{equation}
	
	Before proceeding with the proof of item 1), a positive definite storage function $\mathcal{L}_2(\xi_t,v)$ will be defined and one will show that if $d_t(t)$ remains positive, $\dot{\mathcal{L}}_2$ is negative semi-definite, which implies that the solutions remain bounded for all $t\geq 0$. Define $\mathcal{L}_2$ as
	\begin{align}
	\mathcal{L}_2(\xi_t,v) =\mathcal{L}_1(\xi_t) + \frac12 v^\top {K_p^t}^{-1} v
	\label{eq:L2_1}
	\end{align}
	where $\mathcal{L}_1(\xi_t)$ is the radially unbounded function given by
	\begin{align}\label{L1}
	\mathcal{L}_1(\xi_t) = \frac 1 {n_t} \sum_{i=1}^{n_t}(\|P_{i}^t(\xi_t)\| - \|P_{i}^t(0)\|).
	\end{align}
	To show that $\mathcal{L}_1(\xi_t)$ is a positive definite function, note that
	\begin{align}
	\frac{\partial\mathcal{L}_1}{\partial \xi_t} &= q_t^\top \\
	\frac{\partial^2\mathcal{L}_1}{\partial \xi_t^{2}} &= Q
	\end{align}
	where $Q = \frac 1 {n_t} \sum_{i=1}^{n_t}\frac{1}{\|P_{i}^t\|}\pi_{p_{i}^t}$ is positive definite, as long as at least two of the vectors $p_i^t$ are non-collinear. It follows that $\mathcal{L}_1$ is a convex function of $\xi_t$, with a global minimum attained when $\frac{\partial\mathcal{L}_1}{\partial \xi_t} = q_t^\top = 0$, or equivalently, when $\xi_t=0$. Since $\mathcal{L}_1(0) = 0$ is the global minimum of the function, one concludes that $\mathcal{L}_1(\xi_t)$ is positive-definite.
	Noting that,
	\begin{align}\label{dL1}
	\dot{\mathcal{L}}_1=q_t^\top v,
	\end{align}
	it follows that
	\begin{align}
	\dot{\mathcal{L}}_2=-\frac{1}{d_t}v^\top{K^t_p}^{-1}K^t_dv \label{eq:dotL2}
	\end{align}
	which is negative semi-definite as long as $d_t$ remains positive and implies that the states $\xi_t(t)$ and $v(t)$ remain bounded for all $t\geq0$. The next steps of the proof consist in proving first Item (1) and then the uniform continuity of \eqref{eq:dotL2} along every system's solution in order to deduce, by application of Barbalat's Lemma, the asymptotic convergence of $v$ to zero and from there one deduces the asymptotic convergence of $\dot{v}$ and then $\xi_t$ to zero (Item 2).
	
	\textit{Proof of Item 1:}	
	Using \eqref{eq:sys-t} and the fact that $d_t(t)=-\eta_t^{\top} \xi_t$ and $\dot{d_t}=-\eta_t{^\top } v$ yields
	\begin{align}
	\ddot d_t=-k_{d_3}^t\frac{\dot d_t}{d_t }-k_{p_3}^t\beta_t \label{eq:ddot d_t}
	\end{align}
	with
	\begin{equation}\scalebox{0.95}{$
		\beta_t(t)= -\eta_t^{\top} q_t=\frac 1 {n_t} \sum_{i=1}^{n_t}\frac{d_t}{\|P_t^i\|}>0,\ \forall t.
		\label{eq:betat}$}
	\end{equation}
	This relation is of course valid as long as $d_t(t)>0$.  From there, direct application of \cite[Th. 1-(2)]{rosa2014optical} shows that if $d_t(0) \in \mathbb{R}^+$, the solution $(d_t,\dot{d}_t) \in (\mathbb{R}^+, \mathbb{R})$ exists and uniformly bounded $\forall t$ and converges asymptotically to $(0,0)$.	
	
	\textit{Proof of Item 2:}
	To show that
	\[\ddot{\mathcal L}_2=-\frac{2}{d_t}v^\top{K^t_p}^{-1}K^t_d \dot{v} + \frac{\dot{d_t}}{d_t}\frac{1}{d_t}v^\top{K^t_p}^{-1}K^t_d v \]
	is bounded and hence $\dot{\mathcal L}_2$ \eqref{eq:dotL2} is uniformly continuous, it suffices to show that  $\frac{\|v\|}{d_t}$ is bounded (so is $\frac{\dot d_t}{d_t}$). For that purposes, consider the dynamics of $\frac{v}{d_t}:$
	\begin{equation}\label{eq:dotv/dt}
	\frac{d}{dt}(\frac{v}{d_t}) = -\frac{1}{d_t}((K^t_d+\dot d_t I) \frac{v}{d_t} + K^t_p q_t).
	\end{equation}
	Since $\dot d_t$ converges asymptotically to zero and $q_t$ is bounded then, by direct application of \cite[Lemma 4]{rosa2014optical} one ensures that $\frac{v}{d_t}$ is bounded.  From there one concludes that $\dot{\mathcal L}_2$ is uniformly continuous and hence $v$
	converges asymptotically to zero.
	
	To prove that $q_t(t)$ (or equivalently $\xi_t$) is asymptotically converging to zero one has to show first $\dot{v}$ is converging to zero.
	From \eqref{eq:sys-t}, one can verify that:
	\begin{equation}\label{eq:ddotv}
	\ddot v=-\frac{K_d^t}{d_t} \dot{v} +\delta_{\dot v}^0,
	\end{equation}
	with $\delta_{\dot{v}}^0= K_d^t \frac{\dot d_t}{d_t} \frac{v}{d_t}-K_p^t\dot{q}_t$.
	Since $\frac{v}{d_t}$ (and hence $\frac{\dot d_t}{d_t}$) is bounded and
	\[\dot q_t = Q v=Q_0 \frac{v}{d_t}, \mbox{ with } Q_0=\frac1{n_t} \sum_{i=1}^{n_t} \frac{d_t}{\|P_i^t\|} \pi_{p_i^t}<I_3,\]
	is also a bounded vector, one ensures that $\delta_{\dot{v}}^0$ is bounded. Therefore, direct application of \cite[Lem.~3]{rosa2014optical} concludes boundedness and the asymptotic convergence of $\dot{v}$ to zero and hence one has:
	\begin{equation}
	\frac{v}{d_t} =-{K_d^t}^{-1} K_p^t q_t +o(t)
	\label{eq:lemma5}
	\end{equation}
	with $o(t)$ a asymptotically vanishing term.
	
	By multiplying both sides of the above equation by the bounded vector $q_t^\top$ (the gradient of ${\mathcal L}_1$) and using the fact that $\dot {\mathcal L}_1=q_t^\top v$ \eqref{dL1}, one obtains:
	\begin{align}
	\dot{\mathcal{L}}_1&=  -d_tq_t^\top{K_d^t}^{-1} K_p^t q_t+d_tq_t^\top o(t). \label{eq:dotL1TH1}
	\end{align}
	Since $d_t(t)$ converges asymptotically to zero, then by taking the integral of (\ref{eq:ddot d_t})
	\begin{align}
	\dot{d_t}(t)-\dot{d_t}(0)&=-k^t_{d3} \log(\frac{d_t(t)}{d_t(0)} )- k^t_{p3} \int_{0}^{t}\beta_t(\tau)d\tau, \label{eq:intdt}
	\end{align}
	one concludes that
	\begin{align}
	\lim_{t\rightarrow\infty}\int_0^t \beta_t(\tau) d\tau = +\infty.\label{eq:int_d}
	\end{align}
	Combining equation \eqref{eq:int_d} with the fact that $d(t)\geq \beta_t(t)$ (from \eqref{eq:betat}) and replacing the time index $t$ of equation \eqref{eq:dotL1TH1} by the new time-scale
	index $s(t) :=\int_0^t d_t(\tau) d\tau$ ($s$ tends to infinity if
	and only if $t$ tends to infinity), one has:
	\begin{equation}\label{eq:dsdL}
	\frac{d}{ds}\mathcal{L}_1=  -q_t^\top{K_d^t}^{-1} K_p^t q_t +q_t^\top o(t),
	\end{equation}
	from which one concludes that $q_t$ (and $\xi_t$) is asymptotically converging to zero.
\end{proof}
\subsection{Proof of Proposition \ref{prop:landing}}
\begin{proof}
The proof follows and exploits the same technical steps of the proof of Theorem 1. Since assertions made are almost the same using either \eqref{eq:FT2} or \eqref{eq:FT} (equivalently \eqref{eq:FT2} with $\phi^*_t=0$) except for the last item iii), the proof will be provided using $\eqref{eq:FT2}$ as feedback control and differences will be specified when necessaries.

When $\triangle \neq 0$ and $\phi^*_t \neq 0$ , it is straightforward to verify that \eqref{eq:dotL2} becomes:
\begin{align}
	\dot{\mathcal{L}}_2&=-\frac{1}{d_t}v^\top{K_p^t}^{-1}K_d^tv+v^\top{K_p^t}^{-1}(\triangle + K_d^t \phi^*_t\eta_t)\label{eq:Lyap-dp}
\end{align}
Recall now the dynamics of $\dot{d}_t$ \eqref{eq:ddot d_t}, $\frac{v}{d_t}$ \eqref{eq:dotv/dt}, and of $\ddot{v}$ \eqref{eq:ddotv} in case where $\triangle \neq 0$ and $\phi^*_t \neq 0$.
	\begin{align}
	\ddot d_t &=-k_{d_3}^t\frac{\dot d_t}{d_t }-k_{p_3}^t \beta_t^\triangle \label{eq:ddot d_tp}\\
    \frac{d}{dt}(\frac{v}{d_t}) &= -\frac{1}{d_t}((K_d^t+\dot d_t I) \frac{v}{d_t} + \delta_v^\triangle) \label{eq:dv/dp} \\
    \ddot v &=-\frac{K_d^t}{d_t} \dot{v} +\delta_{\dot v}^\triangle  \label{ddvp},
    \end{align}
with
  \begin{align}
		\beta_t^\triangle(t)&= \frac{1}{k_{p_3}^t}(k_{d_3}^t\phi_t^* +\eta_t^\top \triangle) -\eta_t^{\top} q_t \label{eq:alpha}\\
 \delta_v^\triangle &=K_p^t q_t-\triangle -K_d^t \eta_t\phi_t^*\label{deltav}\\
        \delta_{\dot{v}}^\triangle &= K_d \frac{\dot d_t}{d_t} \frac{v}{d_t}-K_p^t\dot{q}_t+ \dot{\triangle}. \label{deltadv}
  \end{align}
Now since $\beta_t^\triangle(t) >0, \forall t$ independently from the value chosen for $\phi_t^*$, direct application of \cite[Th. 1-(2)]{rosa2014optical} shows that the solution $(d_t,\dot{d}_t) \in (\mathbb{R}^+, \mathbb{R})$ exists and uniformly bounded $\forall t$ and converges (at least) asymptotically to $(0,0)$.

By combining this with the fact that all terms involved in $\delta_v^\triangle$ ($q_t$, $\triangle$ and $\phi^*_t$) are bounded, direct application of \cite[Lem.~4]{rosa2014optical} concludes the boundedness of $\frac{v}{d_t}$. Since $d_t$ is converging to zero, one concludes that $v$ is converging to zero by a direct application of \cite[Lem.~3]{rosa2014optical}. Using the fact that $\dot \triangle$ is bounded by assumption, the proof of boundedness $\ddot{v}$ \eqref{ddvp} and its convergence to zero is directly deduced from to proof the unperturbed case \eqref{eq:ddotv}.  From there and analogously to the unperturbed case (Theorem 1- proof of Item 2), one gets:
\begin{equation}
	\frac{v}{d_t} =-{K_d^t}^{-1} K_p^t q_t+{K_d^t}^{-1}\triangle +\eta_t\phi_t^* +o(t)
	\label{eq:lemma5p}
	\end{equation}
with $o(t)$ an asymptotically vanishing term.

By multiplying both sides of \eqref{eq:lemma5p} by $q_t^\top$ and using the fact that $\dot {\mathcal L}_1=q_t^\top v$ \eqref{dL1}, one obtains:
	\begin{align}
	\frac{\dot{\mathcal{L}}_1}{d_t}&=  -q_t^\top{K_d^t}^{-1} K_p^t q_t +q_t^\top({K_d^t}^{-1}\triangle +\eta_t\phi_t^* +o(t)). \label{eq:dotL1}
	\end{align}

From there one distinguishes between the two issues stated in the proposition:\\
\underline{1) $\eta_t^\top \triangle(t) =0, \forall t$ and $\phi_t^*=0$ ($F_t$ given by \eqref{eq:FT})}\\
By changing the time scale index and similarly to argument used at the end of the proof of Theorem \ref{th:landing}, one concludes that  $\|q_t\|$ is ultimately bounded by $\frac{\|\triangle\|_{\max}}{k_{p_{1,2}}^t}$. Since $d_t=\eta_t^\top \xi_t$ converges to zero, one concludes that $\|\pi_{\eta_t}\xi_t\|$ is ultimately bounded by $\Delta_\xi$ which is the solution of $\|\pi_{\eta_t} q_t \|= \frac{\|\triangle\|_{\max}}{k^t_{d_{1,2}}}$.  \\
\underline{2) $\eta_t^\top \triangle(t) \neq 0$, and $\phi_t^*\neq 0$ ($F_t$ given by \eqref{eq:FT2})}\\
In that case one concludes that the storage function ${\mathcal L}_1$ is decreasing as long as the right hand side of the above equation is negative and $d_t>0$ and hence $\xi_t$ is bounded. The argument of changing the time index is not valid in this case.
\end{proof}

\subsection{Proof of Proposition \ref{prop:window}}
\begin{proof}
	We will consider hereafter only the case where $\sigma(q_w)=1$ (or equivalently when $\eta_w^\top q_w <0$). That is the situation in which the vehicle is going through the window while avoiding collision with the wall and the window edges.
	
	From the dynamics of the closed-loop
	system \eqref{eq:closedloop_w}, the proof focus first on the evolution of $d_w$. That is the evolution of the system in the direction $\eta_w$.
	
	When $\|{q}_w(t)\| \ge \epsilon+\delta$, one has $d_w=d_o=-\eta_w^T \xi_w$ and hence:
	\begin{align}
	\dot{d_o}&=-\eta_w^\top v \notag\\
	\ddot{d_o}&=-k^w_{\phi}\frac{\dot{d_o}}{d_{o}} -k^w_{\phi} \beta_w\label{eq:ddot}
	\end{align}
	with $\beta_w=\phi^*_w+\frac{\eta_w^\top \triangle}{k^w_{\phi}} \geq  \epsilon$.
	
	When $\epsilon<\|{q}_w(t)\| < \epsilon+\delta$, one has $d_w=\frac{d_o d_e}{\alpha_w d_e + (1-\alpha_w) d_o}$ with $\alpha_w$ (defined by \eqref{eq:alphaw}) a uniformly continuous and bounded valued function on $[0,1]$, and hence one verifies that:	
	\begin{equation}
	\ddot d_o(t)=-k^w_\phi b(t)\frac {\dot d_o(t)} {d_o(t)}-k^w_\phi \beta_w \label{eq:dotdw}	
	\end{equation}
	with $b(t)=\frac{(1-\alpha_w(t))d_o(t)+\alpha_w(t)d_e(t)}{d_e(t)}$ a positive uniformly continuous and bounded function as long as $\epsilon< \|q_w(t)\| < \epsilon+\delta$ and $d_o(0)\in \mathbb{R}^+$.
	direct application of \cite[Th. 5.1]{herisse12} to both equation \eqref{eq:ddot} and \eqref{eq:dotdw}, one can conclude that as long as $d_o(0)\in \mathbb{R}^+$ and $\|q_w(t)\| > \epsilon$ (or equivalently $\xi_w\not\in {\mathcal W}$), $d_o(t)\in \mathbb{R}^+,\forall t\ge 0$ and $d_o(t)$ converges to zero exponentially (the exponential convergence of $d_o(t)$ is granted due to the fact that $\beta_w \geq \epsilon$)  but never crosses zero and hence the vehicle will never touch the wall in a finite time. Additionally, one also proves, from \cite[Th. 5.1]{herisse12}, that there exists a finite time $t_1\geq 0$ such that $\dot{d}_o(t)<0,\; \forall t \geq t_1$ and hence $d_o$ and $d_w$ are decreasing after $t_1$.
	
	When $\|q_w(t)\|\le \epsilon$ (the situation when $\xi_w \in {\mathcal W}$), one has $d_w=d_e > d_o$. In this case one can easily verify that \eqref{eq:ddot} can be rewritten as:
	\begin{equation} \label{eq:ddot_de}
	\ddot{d}_o=-k(t)\dot{d}_o -k^w_{\phi} \beta_w
	\end{equation}
	with $k(t)=\frac{k^w_\phi}{d_e}$ a upper bounded positive gain as long as $d_e(t)$ is positive. Due to the fact that $\beta_w \geq \epsilon$, $\dot{d}_o$ is ultimately bounded by $-\frac{k^w_\phi \beta_w}{k(t)}\leq -\frac{k^w_\phi \epsilon}{k(t)}$ and hence one immediately ensures that there exists a finite time $t_2 \geq 0$ from which $\dot{d}_o(t) <0, \forall t \geq t_2$. This implies that when $\|q_w(t)\|\le \epsilon\ (\xi_w \in {\mathcal W})$, $d_o$ is decreasing $\forall t \geq t_2$ and hence $d_o$ crosses zero in a finite time $\bar{t} >t_2$. Note that at $t=\bar{t}$, one has $\sigma(q_w(\bar{t}))=0$ according to \eqref{eq:sigma}.
	
	Consider now the the dynamics of the closed-loop system \eqref{eq:closedloop_w} in the plane $\pi_{\eta_w}$. That is the dynamics of $\xi_{\perp}:=\pi_{\eta_w} \xi_w$.  By defining  $v_{\perp}:=\pi_{\eta_w}v$ and $\triangle_{\perp}:=\pi_{\eta_w}\triangle$, one gets:
	
	\begin{align}
	\dot{\xi}_{\perp} =& v_{\perp}\\
	\dot v_{\perp} = &-\frac{k_d^w}{d_w}(v_\perp+\frac{k_p^w}{k_d^w}\xi_{\perp})+\triangle_{\perp}.
	\end{align}
	Define a new state
	\begin{equation}\label{z}
	z=v_{\perp}+\frac{k_p^w}{k_d^w}\xi_{\perp},
	\end{equation}
	and the following  positive definite storage function:
	\begin{equation*}
	\mathcal{L}_3=\frac1 2 \|z\|^2+\frac{1}{2}(\frac{k_p^w}{k_d^w})^2\|\xi_{\perp}\|^2,
	\end{equation*}
	with time derivative
	\begin{equation}
	\begin{aligned}
	\dot{\mathcal{L}}_3&=-(\frac{k_p^w}{k_d^w})^3\|\xi_{\perp}\|^2-(\frac{k_d^w}{d_w}-\frac{k_p^w}{k_d^w})\|z\|^2+z^\top \triangle_{\perp}\\
	&\le -(\frac{k_p^w}{k_d^w})^3\|\xi_{\perp}\|^2-\frac{\|z\|}{{k_d^w}}((\frac{{k_d^w}^2}{d_w}-k_p^w)\|z\|-k_d^w\|\triangle_{\perp}\|), \label{eq:dotL4}
	\end{aligned}
	\end{equation}
	which is negative-definite provided that $0<d_w<\frac{{k_d^w}^2}{k_p^w}$ and $\|z\|\ge \frac{d_wk_d^w\|\triangle_{\perp}\|_{\max}}{{k_d^w}^2-d_w k_p^w}$.\\
	\textit{Proof of Item 1:}\\
	To show there exists a finite time $t_w\geq0$ at which the vehicle enters the region $\mathcal W$ and remains there  as long as $\sigma(q_w)=1$, one proceeds using a proof by contradiction in two steps. 
	
	In the first step, assume that $\xi_w$ is not converging to ${\mathcal W}$ in a finite time $t_w$ and hence $\|q_w(t)\|> \epsilon$, $\forall t$.  In the second one, assume that $\xi_w$ is switching indefinitely between the two regions. \\
	i) Consider the situation for which the initial condition is such that $\|q_w(0)\|> \epsilon$  (outside the region ${\mathcal W}$). Using  the fact that  there exists a finite time instant $t_1$ from which $d_w$ is decreasing and converging to zero but never crosses zero in finite time (see the above discussion), one concludes that $z$ \eqref{z} is exponentially converging to zero and hence:
	\[v_{\perp}=\dot{\xi}_{\perp}=-\frac{k_p^w}{k_d^w}{\xi}_{\perp} + o(t),\]
	with $o(t)$ an exponential vanishing term. This in turn implies that $\xi_{\perp}$ (resp. $v_{\perp}$) is converging to zero exponentially.  Combining this with the fact that $d_w(t)$ (resp. $d_o(t)$) is converging to zero,  one concludes that there exists a finite time $t_w$ at which $\|q_w(t_w)\|< \epsilon$ $(\xi_w(t_w) \in {\mathcal W})$, which contradicts the first part of the assumption.\\
	ii) Consider the situation for which the vehicle is switching indefinitely between the two regions. Since $d_o(t)$ (respectively $d_w$) 
	is decreasing $\forall t \geq \max\{t_1,t_2\}$  for both cases of $\|q_w\|>\epsilon$ and $\|q_w\|\le\epsilon$ with the fact that $(\xi_{\perp},v_{\perp})$ converges exponentially to $(0,0)$ (proof of the step (i)), one concludes that there exists a finite time $t_w\geq0$ at which the vehicle enters the region $\mathcal W$ ($\|q_w(t_w)\|\le \epsilon$), and remains there as long as $\sigma(q_w)=1$, which contradicts the assumption.	

Combining this with the discussion following \eqref{eq:dotdw}, one ensures that there exists $\epsilon_1>0$ such that $d_w(t)\ge d_o(t)>\epsilon_1,\ \forall t<t_w$.\\
	\textit{Proof of Item 2:}\\
	When $\xi_w$ is inside the region ($\|q_w\|\le \epsilon$), one guarantees that ${\mathcal L}_3$ \eqref{eq:dotL4} is decreasing as long as $0<d_w<\frac{{k_d^w}^2}{k_p^w}$ and $\|z\|\ge \frac{d_wk_d^w\|\triangle_{\perp}\|_{\max}}{{k_d^w}^2-d_w k_p^w}$. Now since there exists a time $\bar{t}>t_w$ such that $d_o(\bar{t})=0$, one concludes that $t_{\lim}$ exists and it is equal to $\bar{t}$. 
\end{proof}

\bibliography{bibliography}

@book{burton1978thinking,
	title={Thinking in perspective: critical essays in the study of thought processes},
	author={Burton, Andrew and Radford, John},
	volume={646},
	year={1978},
	publisher={Routledge}
}

@article{bertrand11,
	title = {A hierarchical controller for miniature \{VTOL\} UAVs: Design and stability analysis using singular perturbation theory},
	journal = {Control Engineering Practice},
	volume = {19},
	number = {10},
	pages = {1099 - 1108},
	year = {2011},
	issn = {0967-0661},
	doi = {http://dx.doi.org/10.1016/j.conengprac.2011.05.008},
	url = {},
	author = {Sylvain Bertrand and Nicolas Gu\'{e}nard and Tarek Hamel and H\'{e}l\´{e}ne Piet-Lahanier and Laurent Eck}
}

@incollection{chaumette2016visual,
	title={Visual servoing},
	author={Chaumette, Fran{\c{c}}ois and Hutchinson, Seth and Corke, Peter},
	booktitle={Springer Handbook of Robotics},
	pages={841--866},
	year={2016},
	publisher={Springer}
}

@article{desouza2002vision,
	title={Vision for mobile robot navigation: A survey},
	author={DeSouza, Guilherme N and Kak, Avinash C},
	journal={IEEE transactions on pattern analysis and machine intelligence},
	volume={24},
	number={2},
	pages={237--267},
	year={2002},
	publisher={IEEE}
}

@article{floreano2015science,
	title={Science, technology and the future of small autonomous drones},
	author={Floreano, Dario and Wood, Robert J},
	journal={Nature},
	volume={521},
	number={7553},
	pages={460},
	year={2015},
	publisher={Nature Publishing Group}
}

@inproceedings{falanga2017aggressive,
	title={Aggressive quadrotor flight through narrow gaps with onboard sensing and computing using active vision},
	author={Falanga, Davide and Mueggler, Elias and Faessler, Matthias and Scaramuzza, Davide},
	booktitle={2017 IEEE international conference on robotics and automation (ICRA)},
	pages={5774--5781},
	year={2017},
	organization={IEEE}
}

@article{guo2020image,
	title={Image-based estimation, planning, and control for high-speed flying through multiple openings},
	author={Guo, Dejun and Leang, Kam K},
	journal={The International Journal of Robotics Research},
	pages={0278364920921943},
	year={2020},
	publisher={SAGE Publications Sage UK: London, England}
}

@ARTICLE{hutchinson96, 
	author={Hutchinson, S. and Hager, G.D. and Corke, P.I.}, 
	journal={Robotics and Automation, IEEE Transactions on}, 
	title={A tutorial on visual servo control}, 
	year={1996}, 
	month={Oct}, 
	volume={12}, 
	number={5}, 
	pages={651-670}, 
	doi={10.1109/70.538972}, 
	ISSN={1042-296X},}

@article{hamel2002dynamic,
	title={Dynamic modelling and configuration stabilization for an X4-flyer.},
	author={Hamel, Tarek and Mahony, Robert and Lozano, Rogelio and Ostrowski, James},
	journal={IFAC Proceedings Volumes},
	volume={35},
	number={1},
	pages={217--222},
	year={2002},
	publisher={Elsevier}
}

@ARTICLE{hamel02, 
	author={Hamel, T. and Mahony, R.}, 
	journal={IEEE Transactions on Robotics and Automation}, 
	title={Visual servoing of an under-actuated dynamic rigid-body system: an image-based approach}, 
	year={2002}, 
	month={Apr}, 
	volume={18}, 
	number={2}, 
	pages={187-198}, 
	doi={10.1109/TRA.2002.999647}, 
	ISSN={1042-296X}
}

@ARTICLE{herisse12, 
	author={H\'{e}riss\'{e}, B. and Hamel, T. and Mahony, R. and Russotto, F.-X.}, 
	journal={IEEE Transactions on Robotics}, 
	title={Landing a VTOL Unmanned Aerial Vehicle on a Moving Platform Using Optical Flow}, 
	year={2012}, 
	month={feb.}, 
	volume={28}, 
	number={1}, 
	pages={77-89}, 
	doi={10.1109/TRO.2011.2163435}, 
	ISSN={1552-3098}
}

@article{ho2018optical,
	title={Optical-flow based self-supervised learning of obstacle appearance applied to mav landing},
	author={Ho, HW and De Wagter, C and Remes, BDW and De Croon, GCHE},
	journal={Robotics and Autonomous Systems},
	volume={100},
	pages={78--94},
	year={2018},
	publisher={Elsevier}
}

@article{lebras13,
	author = {Le Bras, Florent and Hamel, Tarek and Mahony, Robert and Barat, Christian and Thadasack, Julien},
	title = {Approach maneuvers for autonomous landing using visual servo control},
	journal={IEEE Transactions on Aerospace and Electronic Systems},
	year = {2014},
	volume={50}, 
	number={2},
	pages={1051-1065}
}

@article{loianno2017estimation,
	title={Estimation, control, and planning for aggressive flight with a small quadrotor with a single camera and IMU},
	author={Loianno, Giuseppe and Brunner, Chris and McGrath, Gary and Kumar, Vijay},
	journal={IEEE Robotics and Automation Letters},
	volume={2},
	number={2},
	pages={404--411},
	year={2017},
	publisher={IEEE}
}

@ARTICLE{mahony05,
	title={Image-based visual servo control of aerial robotic systems using linear image features},
	author={Mahony, R. and Hamel, T.},
	journal={IEEE Transactions on Robotics},
	year={2005},
	month={April},
	volume={21},
	number={2},
	pages={227-239},
	doi={10.1109/TRO.2004.835446},
	ISSN={1552-3098}
}

@article{mahony08,
	author = {Mahony, R. and Corke, P. and Hamel, T.},
	title = {Dynamic Image-Based Visual Servo Control Using Centroid and Optic Flow Features},
	publisher = {ASME},
	year = {2008},
	journal = {Journal of Dynamic Systems, Measurement, and Control},
	volume = {130},
	number = {1},
	eid = {011005},
	numpages = {12},
	pages = {011005},
	doi = {10.1115/1.2807085}
}

@article{michael2012collaborative,
	title={Collaborative mapping of an earthquake-damaged building via ground and aerial robots},
	author={Michael, Nathan and Shen, Shaojie and Mohta, Kartik and Mulgaonkar, Yash and Kumar, Vijay and Nagatani, Keiji and Okada, Yoshito and Kiribayashi, Seiga and Otake, Kazuki and Yoshida, Kazuya and others},
	journal={Journal of Field Robotics},
	volume={29},
	number={5},
	pages={832--841},
	year={2012},
	publisher={Wiley Online Library}
}

@article{peters1988dynamic,
	title={Dynamic inflow for practical applications},
	author={Peters, David A and HaQuang, Ninh},
	year={1988}
}

@article{rosa2014optical,
	title={Optical-Flow Based Strategies for Landing VTOL UAVs in Cluttered Environments},
	author={Rosa, Lorenzo and Hamel, Tarek and Mahony, Robert and Samson, Claude},
	journal={IFAC Proceedings Volumes},
	volume={47},
	number={3},
	pages={3176--3183},
	year={2014},
	publisher={Elsevier}
}

@ARTICLE{serra15, 
author={Serra, P. and Cunha, R. and Hamel, T. and Silvestre, C. and Le Bras, F.}, 
journal={Control Systems Technology, IEEE Transactions on}, 
title={Nonlinear Image-Based Visual Servo Controller for the Flare Maneuver of Fixed-Wing Aircraft Using Optical Flow}, 
year={2015}, 
month={March}, 
volume={23}, 
number={2}, 
pages={570-583}, 
doi={10.1109/TCST.2014.2330996}, 
ISSN={1063-6536},}

@article{serra2016landing,
	title={Landing of a Quadrotor on a Moving Target Using Dynamic Image-Based Visual Servo Control},
	author={Serra, Pedro and Cunha, Rita and Hamel, Tarek and Cabecinhas, David and Silvestre, Carlos},
	journal={IEEE Transactions on Robotics},
	volume={32},
	number={6},
	pages={1524--1535},
	year={2016},
	publisher={IEEE}
}

@article{serres2017optic,
	title={Optic flow-based collision-free strategies: From insects to robots},
	author={Serres, Julien R and Ruffier, Franck},
	journal={Arthropod structure \& development},
	volume={46},
	number={5},
	pages={703--717},
	year={2017},
	publisher={Elsevier}
}

@article{tang2018aircraft,
	title={Aircraft Landing Using Dynamic Two-Dimensional Image-Based Guidance Control},
	author={Tang, Zhiqi and Cunha, Rita and Hamel, Tarek and Silvestre, Carlos},
	journal={IEEE Transactions on Aerospace and Electronic Systems},
	volume={55},
	number={5},
	pages={2104--2117},
	year={2018},
	publisher={IEEE}
}

@inproceedings{tang2018going,
	title={Going through a window and landing a quadrotor using optical flow},
	author={Tang, Zhiqi and Cunha, Rita and Hamel, Tarek and Silvestre, Carlos},
	booktitle={2018 European Control Conference (ECC)},
	pages={2917--2922},
	year={2018},
	organization={IEEE}
}

@inproceedings{tang2015homing,
	title={Homing on a moving dock for a quadrotor vehicle},
	author={Tang, Zhiqi and Li, Linqian and Serra, Pedro and Cabecinhas, David and Hamel, Tarek and Cunha, Rita and Silvestre, Carlos},
	booktitle={TENCON 2015-2015 IEEE Region 10 Conference},
	pages={1--6},
	year={2015},
	organization={IEEE}
}

@Misc{vicon,
	title = {Motion Capture Systems from Vicon},
	author= {VICON},
	howpublished = {\url{http://www.vicon.com}},
	year = {2014}
}

@INPROCEEDINGS{zingg10, 
	author={Zingg, S. and Scaramuzza, D. and Weiss, S. and Siegwart, R.}, 
	booktitle={IEEE International Conference on Robotics and Automation (ICRA)}, 
	title={MAV navigation through indoor corridors using optical flow}, 
	year={2010}, 
	pages={3361-3368}, 
	doi={10.1109/ROBOT.2010.5509777}, 
	ISSN={1050-4729}
}
	\bibliographystyle{agsm}
%

\end{document}